\documentclass[preprint,12pt]{elsarticle}
\usepackage{amssymb}
\usepackage{physics}
\usepackage{float}
\usepackage{amsmath}
\usepackage{enumerate}
\usepackage{caption}
\usepackage{subcaption}
\usepackage[colorlinks,linkcolor=red,anchorcolor=blue,citecolor=green]{hyperref}
\usepackage{geometry}
\usepackage{mathrsfs}
\usepackage{xcolor}
\usepackage{tikz}
\usepackage{tikz-3dplot}
\usepackage{textcomp}

\journal{Elsevier}

\begin{document}

\begin{frontmatter}

\title{Electromagnetic Flow Control in Hypersonic Rarefied Environment}

\author[a]{Zhigang PU}
\ead{zpuac@connect.ust.hk}
\author[a,b]{Kun XU\corref{cor1}}
\ead{makxu@ust.hk}
\cortext[cor1]{Cooresponding author}

\address[a]{Department of Mathematics, Hong Kong University of Science and Technology, Clear Water Bay, Kowloon, Hong Kong, China}
\address[b]{Shenzhen Research Institute, Hong Kong University of Science and Technology, Shenzhen, China}

\begin{abstract}

The Unified Gas-Kinetic Wave-Particle (UGKWP) method, developed for the multiscale simulation of partially ionized plasmas, is applied to model electromagnetic flows around a hemisphere spanning regimes from near-continuum to rarefied conditions. To the best of our knowledge, this study presents the first application of a multiscale plasma solver to such a problem. In the formulation, neutrals, ions, and electrons are treated as distinct species. The numerical implementation is validated through comparisons with reference solutions for neutral hypersonic flow around a sphere, as well as with experimental data for a Mach 4.75 pre-ionized argon flow. In both cases, UGKWP results show close agreement with the reference and experimental data. A further comparative study across different Knudsen numbers demonstrates that rarefaction effects weaken the influence of electromagnetic control. These findings highlight the capability of UGKWP in modeling electromagnetic control problems and underscore the significant role of rarefied effects in predicting flow-control behavior, thereby emphasizing the necessity of multiscale modeling in plasma flow applications.

\end{abstract}

\begin{keyword}
unified gas-kinetic wave-particle method \sep partially-ionized plasma \sep electromagnetic flow control \sep multiscale algorithm
\end{keyword}

\end{frontmatter}


\section{Introduction}
\label{introduction}

Magnetohydrodynamic (MHD) thermal protection technology holds strategic significance for both national defense systems and deep-space exploration missions \cite{ali2024magnetoaerodynamics}. During hypersonic flight, a strong externally applied magnetic field interacts with the weakly ionized gas produced by intense shock waves, increasing the shock standoff distance and thereby reducing the heat flux to the vehicle surface. The theoretical foundations of this technology date back to the 1950s US-Soviet space race, when Resler and Sears pioneered the formulation of fundamental magneto-aerodynamic equations \cite{resler1958}. They modeled weakly ionized plasma in high-speed, high-enthalpy flows as conducting fluids, incorporating Lorentz force and Joule heating terms into the modified Navier–Stokes equations.

Over subsequent decades, considerable progress was made in theoretical, computational, and experimental aspects of MHD thermal protection. In 1958, Bush provided an analytical solution describing the magnetic field’s effect on shock standoff distance \cite{bush1958}, while Ziemer and Bush later provided experimental validation using a Mach 4.5 sphere model \cite{ziemer1958}, marking important early milestones. Cambel further investigated the Hall effect’s detrimental impact on magnetic control efficiency \cite{cambel1967a} and conducted corresponding experiments using pre-ionized argon gas \cite{cambel1967}, creating benchmark datasets that remain valuable for numerical validation. Following these foundational studies, advances in computational fluid dynamics enabled high-fidelity simulations of MHD flows \cite{poggie2002, fujino2008, kai2017thermal}. Recent numerical studies have increasingly incorporated complex physical phenomena, such as rarefied gas effects and all plasma sheath and wall interactions. For example, Fawley \cite{fawley2022assessment} utilized Direct Simulation Monte Carlo (DSMC) methods to model conductivity in transitional flow regimes, while Katsurayama \cite{katsurayama2008kinetic} applied DSMC to simulate rarefied gas effects, approximating the Lorentz force as a body force rather than explicitly resolving charged particle trajectories. Parent \cite{parent2023} further quantified the impact of plasma sheaths on the overall flow field.

Despite these advancements, most numerical simulations rely on MHD equations, whose applicability diminishes under high-altitude conditions characterized by large Knudsen numbers. In such regimes, the continuum assumption inherent to MHD-based models breaks down \cite{lani2023}. DSMC-based approaches become computationally prohibitive due to the need for excessively dense meshes and small time steps, which are restricted by the mean free path scale \cite{boyd2017nonequilibrium}. Accurately capturing flow physics across continuum, transitional, and rarefied regimes thus requires a genuinely multiscale numerical approach. Addressing these challenges, recent developments have led to the Unified Gas-Kinetic Wave-Particle (UGKWP) method, which enables robust modeling of multiscale neutral gas flows from the rarefied to continuum regimes \cite{liu2020unified, zhu2019unified}. The UGKWP framework has since been systematically extended to a variety of transport phenomena, including plasma \cite{liu2021unified, pu2024gas}, granular flows \cite{yang2022unified, yang2024unified}, radiative transfer \cite{liu2023implicit, yang2025unified}, and phonon transport \cite{liu2025unified}, demonstrating its capability to capture non-equilibrium effects. Most recently, UGKWP has been specifically applied to partially ionized plasma flows \cite{pu2025unified}. In the continuum regime, UGKWP accurately recovers ideal MHD, dissipative MHD, and two-fluid models; as the flow transitions to rarefied conditions, the method seamlessly shifts to a particle-based description, effectively capturing kinetic plasma behavior and recovering Particle-in-Cell (PIC) dynamics in the collisionless limit.

In this work, we apply the UGKWP method to modeling the electromagnetic control problems in the rarefied environments. The remainder of this paper is organized as follows: Section \ref{sUGKWPmethod} provides a concise overview of the UGKWP method for partially ionized plasma. Section \ref{sNumericalResults} presents the numerical test results. Finally, Section \ref{sConclusions} offers the main conclusions of this study.

\section{UGKWP Method}
\label{sUGKWPmethod}

\subsection{Governing equations}
In this subsection, we introduce the physical models and governing equations utilized in this study. Theoretically, the BGK-Maxwell system can be used to model the partially-ionized plasma \cite{pu2025unified},
\begin{equation*}
    \begin{aligned}
    & \frac{\partial f_\alpha}{\partial t}+\boldsymbol{u}_\alpha \cdot \nabla_{\boldsymbol{x}} f_\alpha+ \frac{q_\alpha(\boldsymbol{E}+\boldsymbol{u}_\alpha \times \boldsymbol{B})}{m_\alpha} \cdot \nabla_{\boldsymbol{u}} f_\alpha= Q_{\alpha}, \\
    & \frac{\partial \boldsymbol{B}}{\partial t}+\nabla_{\boldsymbol{x}} \times \boldsymbol{E}=0, \quad
     \frac{\partial \boldsymbol{E}}{\partial t}-c^2 \nabla_{\boldsymbol{x}} \times \boldsymbol{B}=-\frac{1}{\epsilon_0} \boldsymbol{J}, \\
    &\nabla_{\boldsymbol{x}} \cdot \boldsymbol{E}=\frac{\rho_c}{\epsilon_0},\quad \nabla_{\boldsymbol{x}} \cdot \boldsymbol{B}=0,
    \end{aligned}
\end{equation*}
where $f_\alpha = f_\alpha(t, \boldsymbol{x}, \boldsymbol{u})$ is the microscopic distribution function for species $\alpha$ ($\alpha=i$ for ion and $\alpha=e$ for electron, $\alpha=n$ for atom) at space and time $(\boldsymbol{x},t)$ and microscopic translational velocity $\boldsymbol{u}$. $q_\alpha$ is the charge carried by species $\alpha$. $\boldsymbol{E}, \boldsymbol{B}$ is electric and magnetic field strength. $\epsilon_0$ is the permittivity of free space and $c$ is the light speed, $\boldsymbol{J}$ is the electric current density. $\rho_c$ is the electric charge density.  $Q_{\alpha}=\sum_{k\neq \alpha}Q_{\alpha k}(f_\alpha, f_k)$ is the collision operator of species $\alpha$ between species $k$, approximated by Andries-Aoki-Perthame (AAP) model \cite{andries2002consistent},
\begin{equation*}
    Q_\alpha=\frac{g_\alpha^M-f_\alpha}{\tau_\alpha},
\end{equation*}
where $g_\alpha^{M}$ is a Maxwellian distribution,
\begin{equation*}
    g_\alpha^M=\rho_\alpha\left(\frac{m_\alpha}{2 \pi k \bar{T}_\alpha}\right)^{3 / 2} \exp \left(-\frac{m_\alpha}{2 k_B \bar{T}_\alpha}\left(\boldsymbol{u}_\alpha-\bar{\boldsymbol{U}}_\alpha\right)^2\right),
\end{equation*}
and post-collision velocity $\bar{\boldsymbol{U}}_\alpha$ and temperature $\bar{T}_\alpha$ are given as:
\begin{equation*}
    \begin{aligned}
    \bar{\boldsymbol{U}}_\alpha & =\boldsymbol{U}_\alpha+\frac{\tau_\alpha}{m_\alpha} \sum_{k=1}^N 2 \mu_{\alpha k} \chi_{\alpha k} n_k\left(\boldsymbol{U}_k-\boldsymbol{U}_\alpha\right), \\
    \bar{T}_\alpha & =T_\alpha-\frac{m_\alpha}{3 k_B}\left(\bar{\boldsymbol{U}}_\alpha-\boldsymbol{U}_\alpha\right)^2+\tau_\alpha \sum_{k=1}^N \frac{4 \mu_{\alpha k} \chi_{\alpha k} n_k}{m_\alpha+m_k}\left(T_k-T_\alpha+\frac{m_k}{3 k_B}\left(\boldsymbol{U}_k-\boldsymbol{U}_\alpha\right)^2\right),
    \end{aligned}
    \label{eq:aap U and E}
\end{equation*}
where $\mu_{\alpha k} = m_{\alpha}m_k/(m_{\alpha}+m_k)$ is reduced mass, $N$ is number of species in the system. Species $\alpha$'s mean relaxation time  $\tau_\alpha$ is determined by $1 / \tau_\alpha=\sum_{k=1}^m \chi_{\alpha k} n_{k} $, and interaction coefficient $\chi_{\alpha k}$ for variable hard sphere (VHS) model is \cite{morse1963}:
\begin{equation*}
    \chi_{\alpha k}= \frac{(5-2\omega )(7-2\omega)\sqrt{ \pi }}{15}\left(\frac{2 k_B T_{\alpha}}{m_\alpha}+\frac{2 k_B T_{k}}{m_k}\right)^{1 / 2}\left(\frac{d_\alpha+d_k}{2}\right)^2 ,
\end{equation*}
where $ d_\alpha, d_k $ are the diameters of the particles. The temperature modifies the molecular diameter as
$
({d}/{d_{ref}})^2 = \left( {T}/{T_{ref}} \right)^{\frac{1}{2} - \omega},
$
where $\omega$ is the power coefficient between viscosity and temperature \cite{bird1976molecular}.

For hypersonic applications, the BGK-Maxwell system can be simplified. In many such applications, the ionized gas within the shock layer exhibits low electrical conductivity $\sigma^{cond}$ \cite{poggie2002}. This leads to a small induced current density $\boldsymbol{J}$ according to the Ohm's law $\boldsymbol{J} = \sigma^{cond}(\boldsymbol{E}+\boldsymbol{U}\times \boldsymbol{B})$. Consequently, the magnetic field produced by this current is significantly weaker than the externally applied magnetic field $\boldsymbol{B}_{ext}$. This allows for the neglect of magnetic field distortion by the flow field, enabling the assumption of a static magnetic field for the present investigation. Faraday's law can be neglected since the static magnetic field will generate no transverse electric field.
Furthermore, the hyperbolic divergence cleaning technique realizes the charge neutrality \cite{munz2000divergence}:
$$
\epsilon_0\frac{\partial \boldsymbol{E}}{\partial t} + c^2\nabla \phi = 0,\quad \frac{\partial \phi}{\partial t} + \nabla \cdot \boldsymbol{E} = \frac{\rho_c}{\epsilon_0},
$$
where $\phi$ is the artificial correction potential. An approximate Debye length of $10^{-4}$ m is implemented to alleviate numerical stiffness. Therefore, the final form of the physical model is given as,
\begin{equation}
    \begin{aligned}
    & \frac{\partial f_\alpha}{\partial t}+\boldsymbol{u}_\alpha \cdot \nabla_{\boldsymbol{x}} f_\alpha+ \frac{q_\alpha(\boldsymbol{E}+\boldsymbol{u}_\alpha \times \boldsymbol{B}_{ext})}{m_\alpha} \cdot \nabla_{\boldsymbol{u}} f_\alpha= Q_{\alpha}, \\
    &\epsilon_0\frac{\partial \boldsymbol{E}}{\partial t} + c^2\nabla \phi = -\boldsymbol{J},\quad \frac{\partial \phi}{\partial t} + \nabla \cdot \boldsymbol{E} = \frac{\rho_c}{\epsilon_0}.
    \end{aligned}
    \label{eq:final_form}
\end{equation}
In the fluid limit $\tau_{\alpha} \rightarrow 0$, Eq.\eqref{eq:final_form} turns to a multi-fluid equation as
\begin{equation*}
\begin{aligned}
& \partial_t \rho_n+\nabla_{\boldsymbol{x}} \cdot\left(\rho_n \boldsymbol{U}_n\right)=0, \\
& \partial_t\left(\rho_n \boldsymbol{U}_n\right)+\nabla_{\boldsymbol{x}} \cdot\left(\rho_n \boldsymbol{U}_n \boldsymbol{U}_n+p_n \mathbb{I}\right)=S_n, \\
& \partial_t \mathscr{E}_n+\nabla_{\boldsymbol{x}} \cdot\left(\left(\mathscr{E}_n+p_n\right) \boldsymbol{U}_n\right)=Q_n,\\
& \partial_t \rho_i+\nabla_{\boldsymbol{x}} \cdot\left(\rho_i \boldsymbol{U}_i\right)=0, \\
& \partial_t\left(\rho_i \boldsymbol{U}_i\right)+\nabla_{\boldsymbol{x}} \cdot\left(\rho_i \boldsymbol{U}_i \boldsymbol{U}_i+p_i \mathbb{I}\right)=q_in_i\left(\boldsymbol{E}+\boldsymbol{U}_i \times \boldsymbol{B}_{ext}\right)+S_i, \\
& \partial_t \mathscr{E}_i+\nabla_{\boldsymbol{x}} \cdot\left(\left(\mathscr{E}_i+p_i\right) \boldsymbol{U}_i\right)=q_in_i \boldsymbol{U}_i \cdot \boldsymbol{E}+Q_i,\\
& \partial_t \rho_e+\nabla_{\boldsymbol{x}} \cdot\left(\rho_e \boldsymbol{U}_e\right)=0, \\
& \partial_t\left(\rho_e \boldsymbol{U}_e\right)+\nabla_{\boldsymbol{x}} \cdot\left(\rho_e \boldsymbol{U}_e \boldsymbol{U}_e+p_e \mathbb{I}\right)=q_en_e\left(\boldsymbol{E}+\boldsymbol{U}_e \times \boldsymbol{B}_{ext}\right)+S_e, \\
& \partial_t \mathscr{E}_e+\nabla_{\boldsymbol{x}} \cdot\left(\left(\mathscr{E}_e+p_e\right) \boldsymbol{U}_e\right)=q_en_e \boldsymbol{U}_e \cdot\boldsymbol{E}+Q_e, \\
 &\epsilon_0\frac{\partial \boldsymbol{E}}{\partial t} + c^2\nabla \phi = -\boldsymbol{J},\quad \frac{\partial \phi}{\partial t} + \nabla \cdot \boldsymbol{E} = \frac{\rho_c}{\epsilon_0},
\end{aligned}
\end{equation*}
where $\rho_\alpha, \boldsymbol{U}_\alpha, \mathscr{E}_\alpha$ represent the mass density, macroscopic velocity, and energy density of species $\alpha$, and $p_\alpha$ is the thermal pressure of species $\alpha$. $S_\alpha, Q_\alpha$ are the source terms due to cross-species momentum and energy transfer. In the limit of a strong magnetic field, negligible electron mass, and the non-relativistic regime, Eq.\eqref{eq:final_form} simplifies to MHD equations. A detailed asymptotic analysis of this reduction is provided in \cite{pu2025unified}.

\subsection{UGKWP method}
In this subsection, the UGKWP method to solve the system \eqref{eq:final_form} as it transitions from the rarefied regime to the continuum regime is introduced. First, the multispecies collision operator is rewritten as
\begin{equation}
Q_\alpha = \frac{g_\alpha^{eq}-f_\alpha}{\tau_\alpha} + \frac{g_\alpha^M-g_\alpha^{eq}}{\tau_\alpha},
\label{eq:aap-split}
\end{equation}
where $g^{eq}_\alpha$ is the Maxwellian distribution by collision within the same species, and $g^M_\alpha$ is the Maxwellian distribution after collisions between different species.  Then the operator splitting method is used to solve system \eqref{eq:final_form} and \eqref{eq:aap-split} with timestep $\Delta t$,
\begin{align}
    &\mathcal{L}_1: \frac{\partial f_\alpha}{\partial t}+\boldsymbol{u}_\alpha \cdot \nabla_{\boldsymbol{x}} f_\alpha = \frac{g_\alpha^{eq}-f_\alpha}{\tau_\alpha},\label{eq:l1}\\
    &\mathcal{L}_2: \frac{\partial f_\alpha}{\partial t}+ \frac{q_\alpha(\boldsymbol{E}+\boldsymbol{u}_\alpha \times \boldsymbol{B}_{ext})}{m_\alpha} \cdot \nabla_{\boldsymbol{u}} f_\alpha = 0,\quad
    \epsilon_0\frac{\partial \boldsymbol{E}}{\partial t} = -\boldsymbol{J},\label{eq:l2}\\
    &\mathcal{L}_3: \frac{\partial f_\alpha}{\partial t} =  \frac{g_\alpha^M-g_\alpha^{eq}}{\tau_\alpha},\label{eq:l3}\\
    &\mathcal{L}_4:\epsilon_0\frac{\partial \boldsymbol{E}}{\partial t} + c^2\nabla \phi = 0,\quad \frac{\partial \phi}{\partial t} + \nabla \cdot \boldsymbol{E} = \frac{\rho_c}{\epsilon_0}.\label{eq:l4}
\end{align}
The variables after each update are denoted as
\begin{alignat*}{2}
    &\mathcal{L}_1: f^{n}_\alpha \rightarrow f^{*}_\alpha, \boldsymbol{W}^{n}_\alpha \rightarrow \boldsymbol{W}^{*}_\alpha, \quad\quad\;\;
    \mathcal{L}_2: f^{*}_\alpha \rightarrow f^{**}_\alpha, \boldsymbol{W}^{*}_\alpha \rightarrow \boldsymbol{W}^{**}_\alpha, \boldsymbol{E}^{n} \rightarrow \boldsymbol{E}^{*}, \\
    &\mathcal{L}_3:  f^{**}_\alpha \rightarrow f^{n+1}_\alpha, \boldsymbol{W}^{**}_\alpha \rightarrow \boldsymbol{W}^{n+1}_\alpha,\;
    \mathcal{L}_4: \boldsymbol{E}^{*} \rightarrow \boldsymbol{E}^{n+1},\phi^{n} \rightarrow \phi^{n+1}.
\end{alignat*}
where $\boldsymbol{W}_{\alpha} = (\rho_{\alpha}, \rho_{\alpha} \boldsymbol{U}_{\alpha}, \rho_{\alpha} \mathscr{E}_{\alpha})$ is mass, momentum and energy densities of species $\alpha$.

First, we introduce algorithms for solving $\mathcal{L}_1$. For simplicity, the species subscript $\alpha$ is omitted here. BGK equation without the force term is written as
$$
     \frac{\partial f}{\partial t}+\boldsymbol{u} \cdot \nabla_x f = \frac{g-f}{\tau},
$$
the time-dependent solution of the distribution function at the interface can be written as,
\begin{equation}
    f\left(\boldsymbol{x},\boldsymbol{u}, \boldsymbol{\xi}, t\right)=\frac{1}{\tau} \int_{0}^{t} g(\boldsymbol{x}^{\prime}, \boldsymbol{u}, \boldsymbol{\xi},t^{'}) e^{-\left(t-t^{\prime}\right) / \tau} \mathrm{d} t^{\prime} + e^{-t / \tau} f_{0}\left(\boldsymbol{x}-\boldsymbol{u} t \right),
    \label{eq:BGKsoln}
\end{equation}
where $\tau$ is the local mean relaxation time. $f_0$ is the initial gas distribution function at $t=0$, and $g$ is equilibrium distribution along the characteristic line $\boldsymbol{x}^{'} = \boldsymbol{x} - \boldsymbol{u} t^{'}$. Expand the equilibrium distribution function by the Taylor series to the second-order accuracy,
\begin{equation}
g^{'}=g+ {\boldsymbol{g}}_x \cdot(\boldsymbol{x}^{'}-\boldsymbol{x})+g_t(t^{'}-t),
\label{eq:taylor g}
\end{equation}
where  $g\equiv g(\boldsymbol{x}, \boldsymbol{u},t),g^{'}\equiv g(\boldsymbol{x}^{\prime}, \boldsymbol{u}, t^{'})$.
Substitute them into Eq.\eqref{eq:BGKsoln}, the numerical multiscale evolution solution for simulation particle can be obtained,
\begin{equation}
f(\boldsymbol{x}, \boldsymbol{u},  t)=\left(1-e^{-t / \tau}\right) g^{+}(\boldsymbol{x}, \boldsymbol{u}, t )+e^{-t / \tau} f_{0}\left(\boldsymbol{x} - \boldsymbol{u}t\right),
\label{eq:multiscale BGK soln}
\end{equation}
where,
$$
g^{+}\left( \boldsymbol{x},\boldsymbol{u}, t \right) = g\left( \boldsymbol{x},\boldsymbol{u} ,t\right) + \left( \frac{te^{- t\text{/}\tau}}{1 - e^{- t\text{/}\tau}} - \tau \right)\boldsymbol{u}\cdot {\boldsymbol{g}}_x\left( \boldsymbol{x},\boldsymbol{u},t \right) +\left( \frac{t}{1 - e^{- t\text{/}\tau}} - \tau \right)g_t\left( \boldsymbol{x},\boldsymbol{u},t \right).
$$
The equation above describes the solution to the BGK equation, where the distribution function $f$ at time $t$ is a combination of the initial distribution function $f_0$ and the Taylor expansion of the equilibrium state $g$. From the particles' perspective, this equation implies that a particle has a probability of $e^{-t/\tau}$ to freely stream during the time interval $[0,t]$, and a probability of $(1 - e^{-t/\tau})$ to collide with other particles. After multiple collisions, the particle distribution will reach the equilibrium state $g^+$, whose degree of freedom can be degenerated.

Based on this physical picture, the UGKWP method represents the velocity distribution function in a micro-macro coupled way. A portion of the distribution is captured analytically through the equilibrium distribution function $g^+$, while the remaining part is represented by stochastic simulation particles $P_k = (m_k, \boldsymbol{x}_k, \boldsymbol{u}_k)$. Here, $m_k$ denotes the mass of simulation particle $P_k$, which corresponds to a cluster of real gas particles of the same species, and $\boldsymbol{x}_k$ and $\boldsymbol{u}_k$ represent the position and velocity of the simulation particle $P_k$, respectively.

According to Eq.\eqref{eq:multiscale BGK soln}, the cumulative distribution function of the particle's free streaming time $t_{f}$ before the collision is given as
$$
F\left( t_{f} < t \right) = \exp\left( - t\text{/}\tau \right),
$$
from which the free stream time $t_{f}$ can be sampled as $t_{f} = - \tau\ln(\eta)$ with $\eta$ a random varible subject to the uniform distribution $\eta \sim U(0,1)$ . For a time step $\Delta t$ , the particles with $t_{f} \geq \Delta t$ will undergo collisionless free streaming, and the particles with $t_{f} < \Delta t$ will experience collisional interactions. The procedure of updating particles in the UGKWP method is
\begin{enumerate}[Step 1:]

\item During the time step, stream each particle $P_k$ for a time period of $\min(\Delta t, t_{f,k})$. Then identify and retain the collisionless particles, while removing the collisional particles. Calculate the free-transport flux across cell interfaces contributed by the particles and accumulate the total conservative quantities of the particles $\boldsymbol{W}_i^p$;

\item After updating the macroscopic conservative variables, calculate the total conservative quantities of the collisional particles $\boldsymbol{W}_i^h$ from the updated conservative quantities $\boldsymbol{W}_i$ as $\boldsymbol{W}_i^h = \boldsymbol{W}_i - \boldsymbol{W}_i^p$;

\item At the end of the time step, rebuild the velocity distribution. Calculate the analytical distribution $g^{+,c}$ and resample the collisionless particles from the distribution $g^{+,f}$ according to the updated conservative quantities, sample the free-streaming time $t_{f,k}$ for each particle $P_k$ from the cumulative distribution function $F(t_f < t) = \exp(-t/\tau)$ $\boldsymbol{W}_i^h$.
\end{enumerate}
Refer to \cite{pu2025unified} for a more detailed description of the procedure. In the procedure outlined above, the algorithm for updating the macroscopic conservative variables is introduced in the following paragraph. The evolution of the microscopic velocity distribution is now solved numerically using particles.

Based on the finite volume framework, the cell-averaged macroscopic conservative variables on a physical cell $\Omega_i$ are updated by the numerical flux through cell interfaces, electromagnetic forces and cross-species collisions:
\begin{equation*}
(\boldsymbol{W}_{\alpha})_i^{n+1} = (\boldsymbol{W}_{\alpha})_i^n - \frac{\Delta t}{|\Omega_i|}\sum_{s\in\partial\Omega_i}|l_s|(\mathscr{F}_{\boldsymbol{W}_\alpha})_s,
\label{eq:FVM discretization}
\end{equation*}
where $l_s\in \partial \Omega_i$ is the cell interface with center $\boldsymbol{x}_s$ and outer unit normal vector $\boldsymbol{n}_{s}$. $|l_s|$ is the area of the cell interface. $(\boldsymbol{S}_{\alpha})_i$ is source term due to electromagnetic force. The numerical flux across interface $(\mathscr{F}_{\boldsymbol{W}_\alpha})_s$ can be evaluated from distribution function at the interface,
\begin{equation*}
(\mathscr{F}_{\boldsymbol{W}_\alpha})_s = \frac{1}{\Delta t}\int_{t^n}^{t^{n+1}}\boldsymbol{u}\cdot \boldsymbol{n}_sf_\alpha(\boldsymbol{x}_{s},\boldsymbol{u},\boldsymbol{\xi},t)\boldsymbol{\Psi} \mathrm{d}\boldsymbol{\Xi}\mathrm{d}t,
\label{eq:FVM macroscopic flux}
\end{equation*}
where $\boldsymbol{\Psi}=(1,\boldsymbol{u},\frac{1}{2}(\boldsymbol{u}^2+\boldsymbol{\xi}^2))$ is the conservative moments of distribution functions with $\boldsymbol{\xi}=(\xi_1,\xi_2,\cdots,\xi_n)$ the internal degree of freedom. $\mathrm{d}\boldsymbol{\Xi}=\mathrm{d}\boldsymbol{u}d\boldsymbol{\xi}$ is the volume element in the phase space. For the cell interface $l_s$, the distribution function $f_\alpha(\boldsymbol{x}_{s},\boldsymbol{u},\boldsymbol{\xi},t)$ cannot be analytically represented in the transitional and collisionless regimes, thus requiring direct tracking of its evolution as introduced in the previous paragraph.
This numerical flux of the macroscopic conservative variable can be split into the equilibrium flux and free streaming flux according to Eq.\eqref{eq:BGKsoln}. The equilibrium flux is
\begin{equation}
(\mathscr{F}_{\boldsymbol{W}})_s^g = \frac{1}{\Delta t}\int_{t^n}^{t^{n+1}}\boldsymbol{u}\cdot \boldsymbol{n}_s
\left[\frac{1}{\tau} \int_{0}^{t} g(\boldsymbol{x}^{\prime}, \boldsymbol{u}, \boldsymbol{\xi},t^{'}) e^{-\left(t-t^{\prime}\right) / \tau} \mathrm{d} t^{\prime}\right]
\boldsymbol{\Psi} \mathrm{d}\boldsymbol{\Xi}\mathrm{d}t,
\label{eq:macroscopic eq flux}
\end{equation}
and the free streaming flux is
\begin{equation}
(\mathscr{F}_{\boldsymbol{W}})_s^f = \frac{1}{\Delta t}\int_{t^n}^{t^{n+1}}\boldsymbol{u}\cdot \boldsymbol{n}_s
\left[e^{-t / \tau} f_{0}\left(\boldsymbol{x}-\boldsymbol{u} t \right)\right]
\boldsymbol{\Psi} \mathrm{d}\boldsymbol{\Xi}\mathrm{d}t.
\label{eq:macroscopic fr flux}
\end{equation}

The equilibrium flux can be calculated directly from the macroscopic conservative variables. Assume in the equation \eqref{eq:FVM macroscopic flux}, $\boldsymbol{x}_{s} = \boldsymbol{0}$ and $t^{n} = 0$, the equilibrium $g$ at the cell interface is obtained by conservation constraint
$$
\int_{}^{}{g\boldsymbol{\Psi}}d\boldsymbol{\Xi} = \int_{\boldsymbol{v} \cdot \boldsymbol{n} > 0}^{}{g^{l}\boldsymbol{\Psi}}d\boldsymbol{\Xi} + \int_{\boldsymbol{v} \cdot \boldsymbol{n} < 0}^{}{g^{r}\boldsymbol{\Psi}}d\boldsymbol{\Xi}.
$$
The spatial and time derivatives can be obtained,
$$
\int_{}^{}{g_x\boldsymbol{\Psi}}d\boldsymbol{\Xi} = \boldsymbol{W}_x
,\quad \int_{}^{}{g_t\boldsymbol{\Psi}}d\boldsymbol{\Xi} = - \int_{}^{}\boldsymbol{u} \cdot  g_x\boldsymbol{\Psi}d\boldsymbol{\Xi},
$$
where $g^{l}$ and $g^{r}$ are the equilibrium distributions according to the reconstructed left and right side conservative variables at cell interface $\boldsymbol{W}^{l},{\ \boldsymbol{W}}^{r}$, and $\boldsymbol{W}_x$ is the reconstructed spatial derivative of conservative variables at cell interface. The van Leer limiter is used to achieve a second-order accuracy in space reconstruction. Substituting the reconstructed equilibrium distribution into the equilibrium flux, we have
$$
(\mathscr{F}_{\boldsymbol{W}})_s^g = \int_{}^{}\boldsymbol{u} \cdot \boldsymbol{n}_{s}\left( C_{1}g_{0} + C_{2}\boldsymbol{u} \cdot {\boldsymbol{g}}_{0x} + C_{3}g_{0t} \right)\Psi d\boldsymbol{\Xi},
$$
where the time integration coefficients are
\begin{align*}
&C_{1} = \Delta t - \tau\left( 1 - e^{- \Delta t\text{/}\tau} \right)
,\\
&C_{2} = 2\tau^2(1 - e^{- \Delta t\text{/}\tau}) - \tau\Delta t -\tau \Delta t  e^{- \Delta t\text{/}\tau},\\
&C_{3} = \frac{\Delta t^2}{2} - \tau\Delta t + \tau^2(1-e^{\Delta t/\tau}).
\end{align*}

Next we consider the free stream flux $(\mathscr{F}_{\boldsymbol{W}})_s^f$ . As stated in the last subsection, the initial distribution is represented partially by an analytical distribution $g_{a}^{+,c}$, and partially by particles, and therefore the free stream flux is also calculated partially from the reconstructed analytical distribution as  $(\mathscr{F}_{\boldsymbol{W}})_s^{f,w}$, and partially from particles as  $(\mathscr{F}_{\boldsymbol{W}})_s^{f,p}$. The initial analytical distribution $g_{\alpha}^{+ ,c}$ is reconstructed as
$$
g_{0}^{+ ,c}\left( \boldsymbol{x},\boldsymbol{u} \right) = g_{0}^{+ ,c} + {\boldsymbol{g}}_{0x}^{+ ,c} \cdot \boldsymbol{x},
$$
which gives
$$
(\mathscr{F}_{\boldsymbol{W}})_s^{f,w} = \int_{}^{}\boldsymbol{u} \cdot \boldsymbol{n}_s\left( C_{4}g_{0}^{+,c} + C_{5}\boldsymbol{u} \cdot {g}_{0x}^{+,c} \right)\boldsymbol{\Psi}d\boldsymbol{\Xi},
$$
where the time integration coefficients are
\begin{align*}
&C_{4} = \tau\left( 1 - e^{- \Delta t\text{/}\tau} \right) - \Delta te^{- \Delta t\text{/}\tau},
,\\
&C_{5} = \tau\Delta t e^{- \Delta t\text{/}\tau} - \tau^{2}\left( 1 - e^{- \Delta t\text{/}\tau} \right) + \frac{\Delta t^{2}}{2}e^{- \Delta t\text{/}\tau}.
\end{align*}
The net particle flux $(\mathscr{F}_{\boldsymbol{W}})_s^{f,p}$ is calculated as
$$
(\mathscr{F}_{\boldsymbol{W}})_s^{f,p}=  \sum_{k \in P_{\partial\Omega_{i}^{+}}}^{}\boldsymbol{W}_{P_{k}} - \sum_{k \in P_{\partial\Omega_{i}^{-}}}^{}\boldsymbol{W}_{P_{k}},
$$
where $\boldsymbol{W}_{P_{k}} = \left( m_{k},m_{k}\boldsymbol{v}_{k},\frac{1}{2}m_{k}\boldsymbol{v}_{\boldsymbol{k,}}^{2} \right),P_{\partial\Omega_{i}^{-}}$
is the index set of the particles streaming out of cell $\Omega_{i}$
during a time step, and $P_{\partial\Omega_{i}^{+}}$ is the
index set of the particles streaming into cell $\Omega_{i}$. Finally, the update formula for conservative variables is
\begin{align}
    \begin{split}
    \boldsymbol{W}_{\boldsymbol{i}}^{*} = \boldsymbol{W}_{\boldsymbol{i}}^{\boldsymbol{n}} -& \sum_{s}^{}{\frac{\Delta t}{\left| \Omega_{i} \right|}\left| l_{s} \right|(\mathscr{F}_{\boldsymbol{W}})_s^{g}} - \sum_{s}^{}{\frac{\Delta t}{\left| \Omega_{i} \right|}\left| l_{s} \right|(\mathscr{F}_{\boldsymbol{W}})_s^{f,w}} + \frac{1}{\left| \Omega_{i} \right|}(\mathscr{F}_{\boldsymbol{W}})_s^{f,p}     \end{split}
\end{align}
The evolution of the macroscopic conservative variables due to species transport and diffusion has been solved by the coupled evolution of the microscopic particle transport and macroscopic fluid dynamics.

Subsequently, we solve the $\mathcal{L}_2$. The moment equation of Eq. \ref{eq:l2}, is first solved implicitly. Particles are then accelerated using the updated $\boldsymbol{E}^{*}$. Further details regarding the moment equation and the implicit linear system can be found in \cite{pu2025unified}.
Next, we solve $\mathcal{L}_3$, the moment equation of Eq.\eqref{eq:l3} is a set of nonlinear ordinary equations:
\begin{align}
    &\frac{\partial \boldsymbol{U}_{\alpha}}{\partial t}  =  \frac{1}{m_{\alpha}}\sum_{k=1}^N 2 \mu_{\alpha k} \chi_{\alpha k} n_k\left(\boldsymbol{U}_k-\boldsymbol{U}_\alpha\right), \\
    &\frac{\partial E_{\alpha}}{\partial t} = \sum_{k=1}^N \frac{2m_{k}\chi_{\alpha k} n_k}{m_{\alpha}+m_{k}} \boldsymbol{U}_{\alpha}\left(\boldsymbol{U}_k-\boldsymbol{U}_\alpha\right)+\frac{3k_{B}}{2m_{\alpha}} \sum_{k=1}^N \frac{4 \mu_{\alpha k} \chi_{\alpha k} n_k}{m_\alpha+m_k}\left(T_k-T_\alpha+\frac{m_k}{3 k_B}\left(\boldsymbol{U}_k-\boldsymbol{U}_\alpha\right)^2\right)
    \label{eq:aap-moment}
\end{align}
Discretize with the backward Euler scheme and solve it with the Newton iteration solver. The details is in \ref{aap-implicit}. Finally, in $\mathcal{L}_4$, Eq.\eqref{eq:l4} is solved by the flux vector splitting method.

\begin{figure}[h!]
    \centering
    \includegraphics[width=0.8\linewidth]{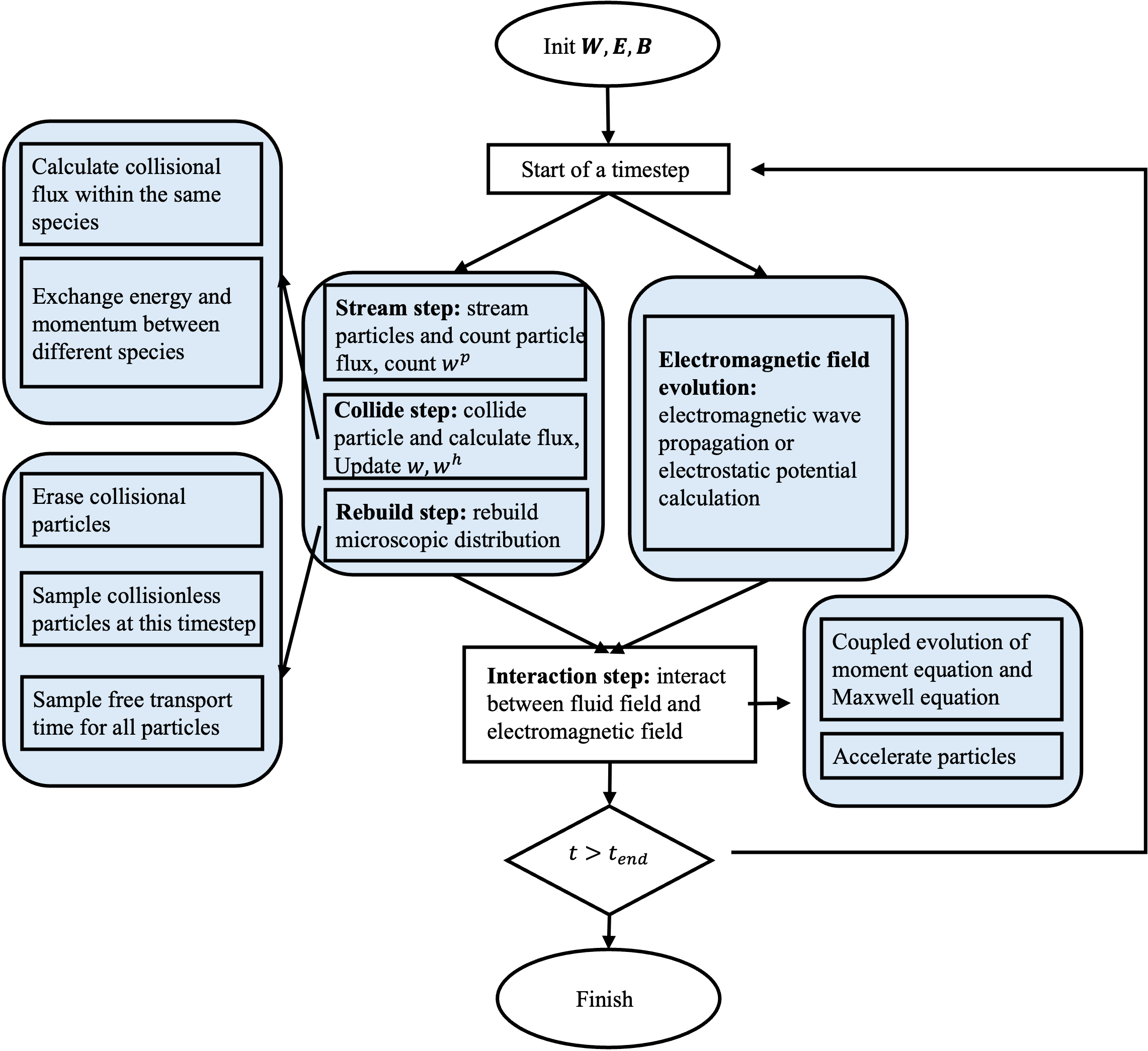}
    \caption{Flow chart of UGKWP method}
    \label{fig:algorithm chart}
\end{figure}

The overall algorithm flow chart is shown in Figure \ref{fig:algorithm chart}.

\section{Numerical Studies}
\label{sNumericalResults}

\subsection{Single Species Hypersonic Nonequilibrium Argon Flow Around a Sphere}

To validate the computational code's accuracy in simulating rarefied gas dynamics under conditions of no ionization, we perform a numerical simulation of a hypersonic neutral argon gas flow around a three-dimensional sphere. This is crucial as an accurate representation of the neutral gas flow field is essential for the subsequent modeling of weakly ionized plasma phenomena. The simulation parameters are set to a Mach number of 4.25 and a Knudsen number of 0.031, with the sphere diameter, $d = 2$ m, serving as the characteristic length. The freestream quantities of argon atoms compute the characteristic Knudsen number. The computational domain is discretized using 132,480 hexahedral cells, with the first layer adjacent to the sphere surface having a height of 0.005$d$. A comprehensive list of simulation parameters is provided in Table \ref{tab:sphere-parameters}. Boundary conditions consist of a far-field inflow and outflow condition at the domain's outer boundaries and an isothermal wall condition imposed on the sphere's surface. The reference number of particles sampled per cell is 400.

\begin{table}[h!]
    \centering
    \begin{tabular}{lc}
        \hline
        \hline
        Parameter & Value \\
        \hline
        Mach number & 4.25 \\
        Knudsen number & 0.031 \\
        Argon atomic mass & $6.63 \times 10^{-26}\ \mathrm{kg}$ \\
        Argon diameter (VHS) & $4.17 \times 10^{-10}\ \mathrm{m}$ \\
        Incoming argon density & $1.3841 \times 10^{-6}\ \mathrm{kg/m^3}$ \\
        Characteristic temperature & 273 K \\
        Wall temperature & 273 K \\
        Incoming argon velocity & 1308.21 m/s \\
        Sphere diameter & 2 m \\
        \hline
    \end{tabular}
    \caption{Simulation parameters for the rarefied argon flow around a sphere.}
    \label{tab:sphere-parameters}
\end{table}

\begin{figure}
    \begin{subfigure}[b]{0.48\textwidth}
    \centering
    \includegraphics[width=0.95\linewidth]{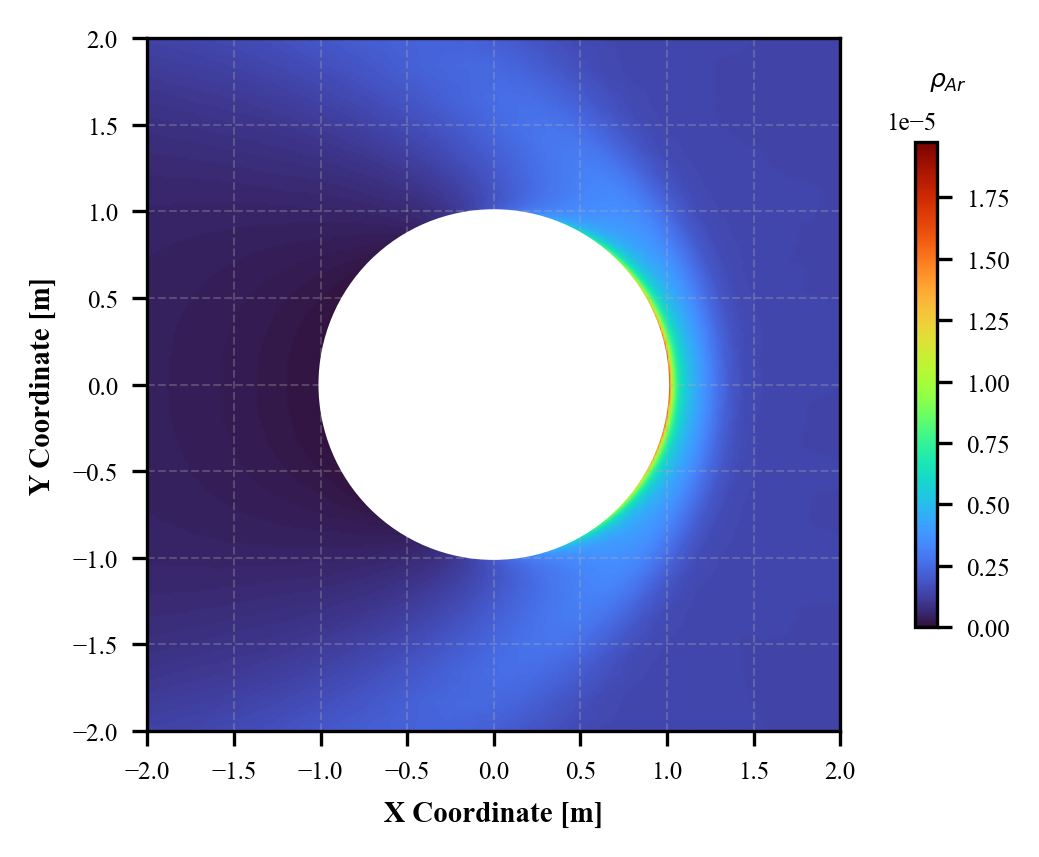}
    \caption{}
    \label{fig:Kn0.031-Density}
    \end{subfigure}
    \hfill
    \begin{subfigure}[b]{0.48\textwidth}
    \centering
    \includegraphics[width=0.95\linewidth]{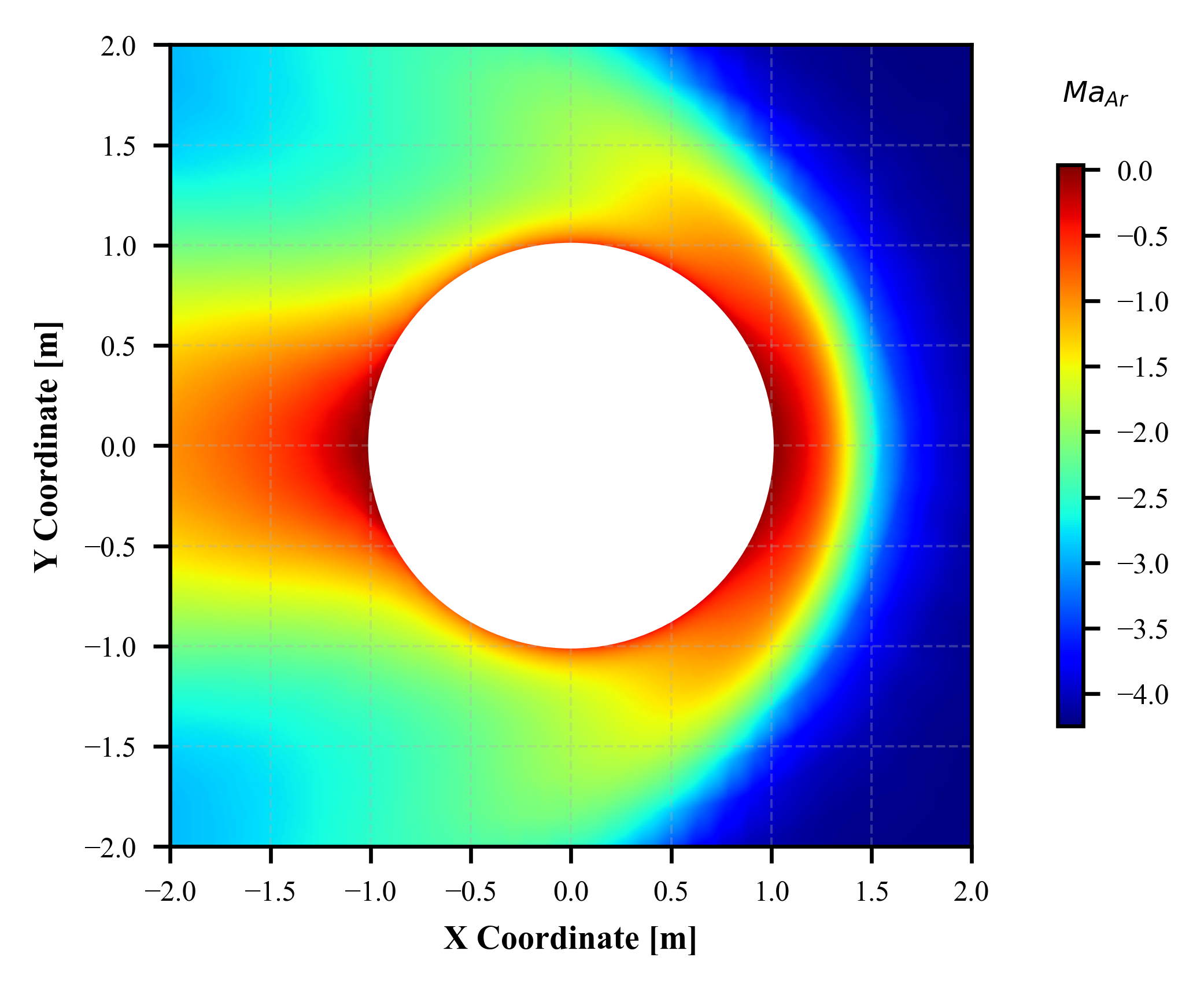}
    \caption{}
    \label{fig:kn0.031-Ma}
    \end{subfigure}
    \vfill
    \begin{subfigure}[b]{0.48\textwidth}
    \centering
    \includegraphics[width=0.95\linewidth]{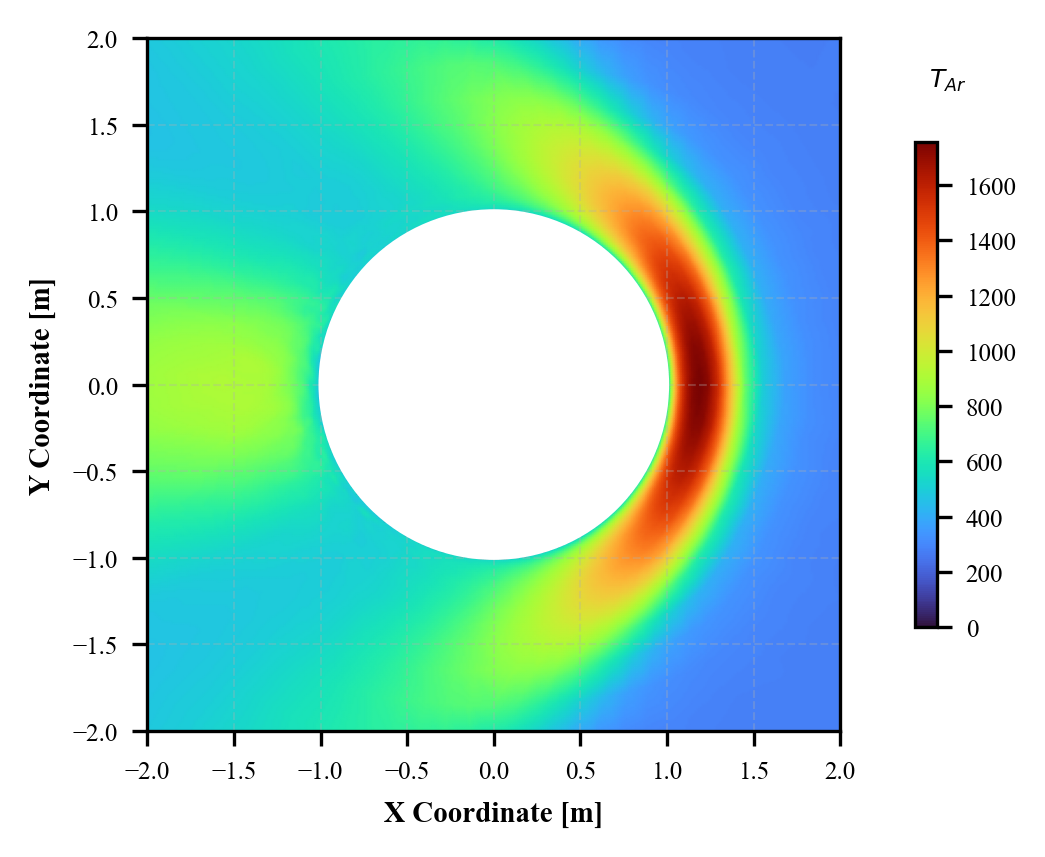}
    \caption{}
    \label{fig:kn0.031-T}
    \end{subfigure}
    \hfill
    \begin{subfigure}[b]{0.48\textwidth}
    \centering
    \includegraphics[width=0.95\linewidth]{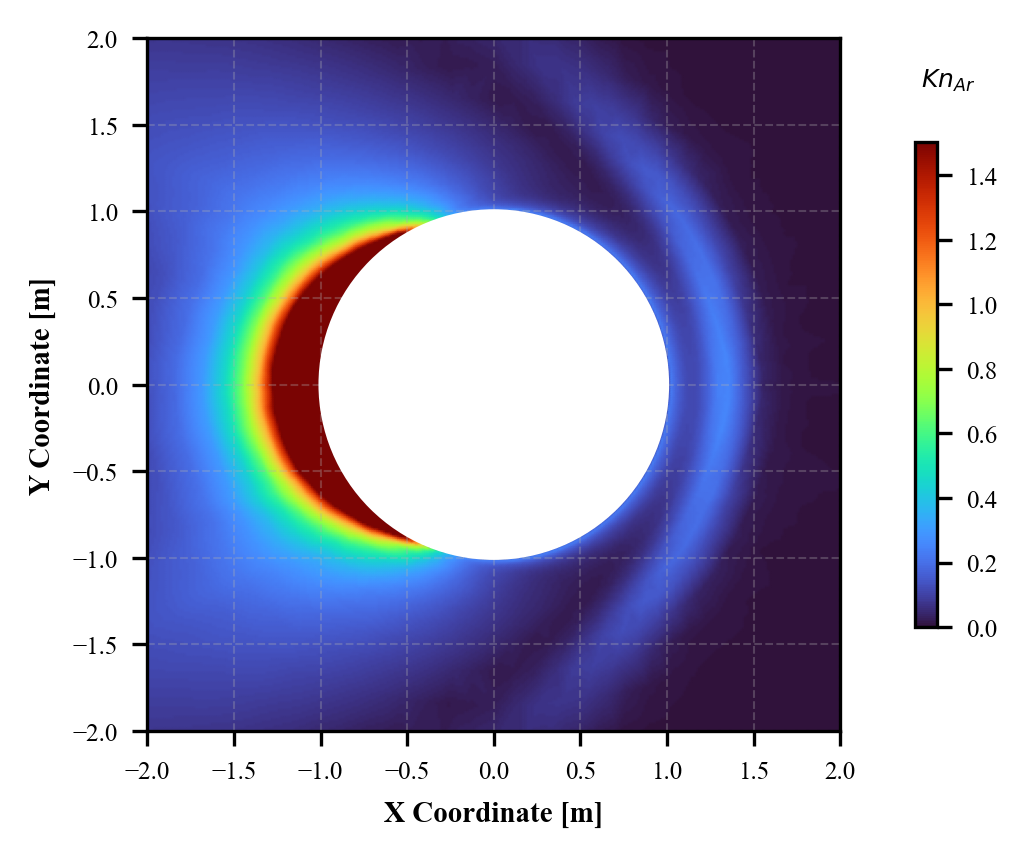}
    \caption{}
    \label{fig:kn0.031-Kn}
    \end{subfigure}
    \caption{ Hypersonic flow around a sphere at Ma=4.25 and Kn=0.031 by the UGKWP method. a) density, b) Mach number, (c) temperature and (d) local Knudsen number.}
    \label{fig:Kn0.031-contour}
\end{figure}

\begin{figure}
    \centering
    \begin{subfigure}[b]{0.3\textwidth}
        \includegraphics[width=\textwidth]{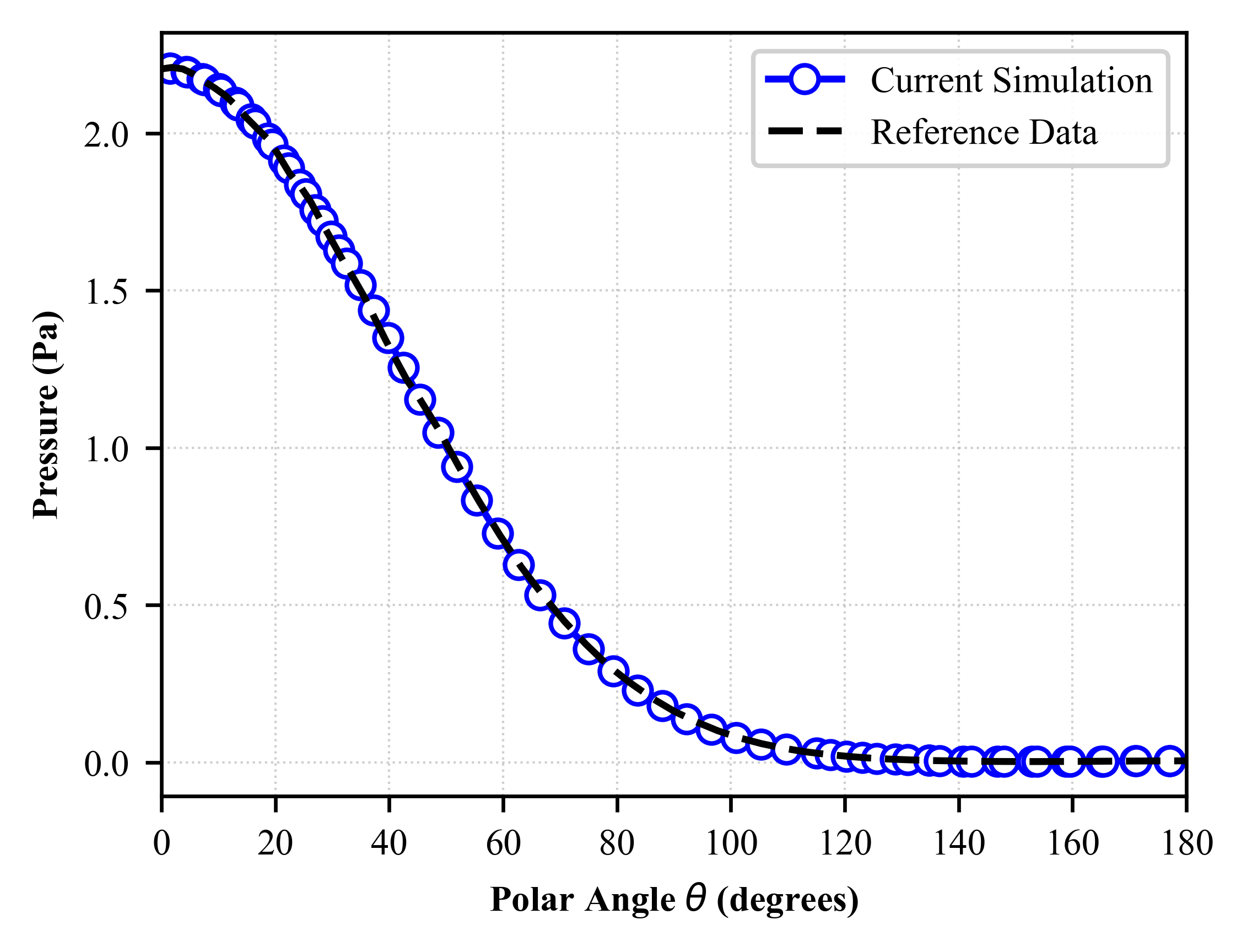}
        \caption{Surface pressure coefficient}
        \label{fig:EXP2M004-pressure_coefficient}
    \end{subfigure}
    \hfill 
    \begin{subfigure}[b]{0.3\textwidth}
        \includegraphics[width=\textwidth]{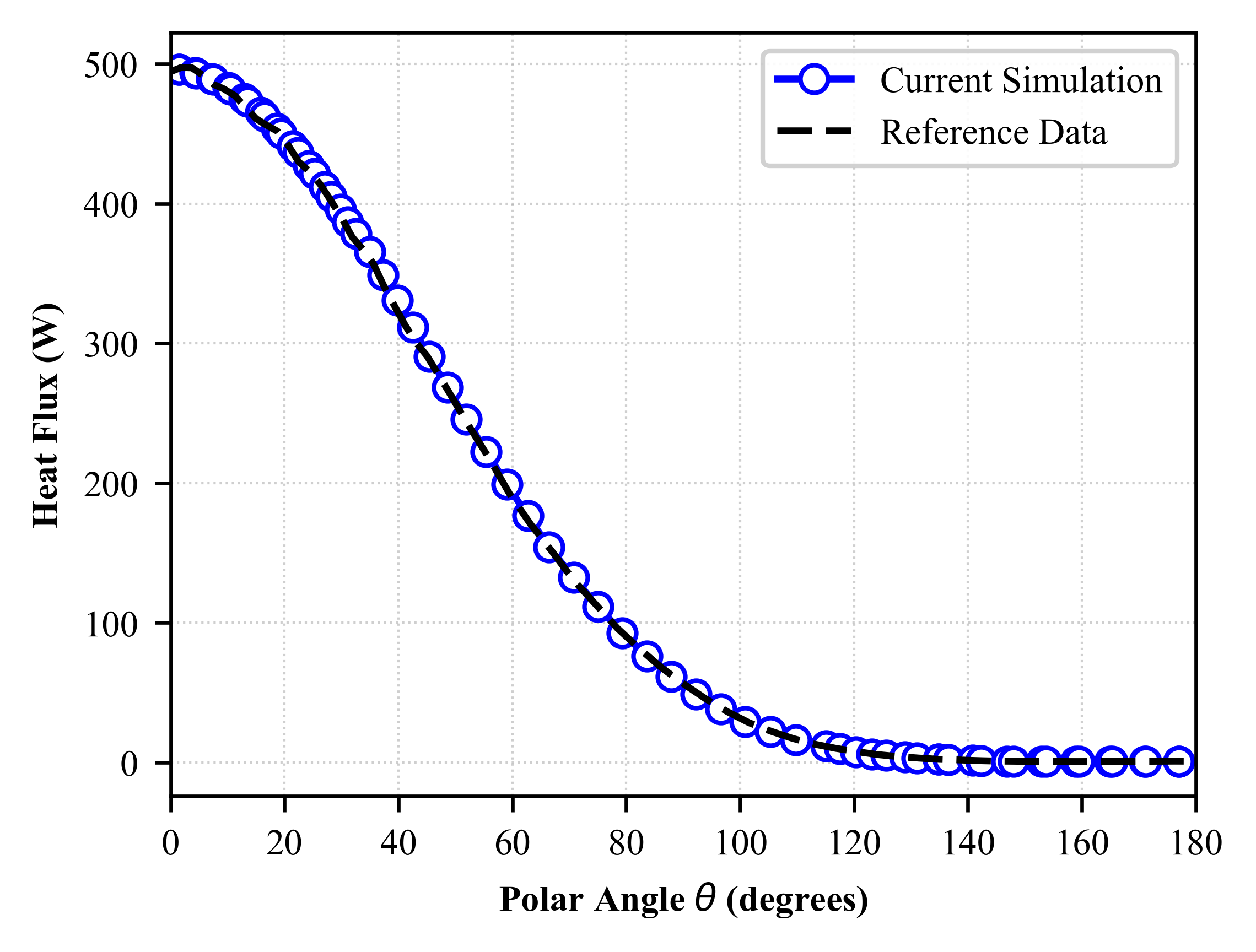}
        \caption{Surface heat flux coefficient}
        \label{fig:EXP2M004-heat_flux_coefficient}
    \end{subfigure}
    \hfill 
    \begin{subfigure}[b]{0.3\textwidth}
        \includegraphics[width=\textwidth]{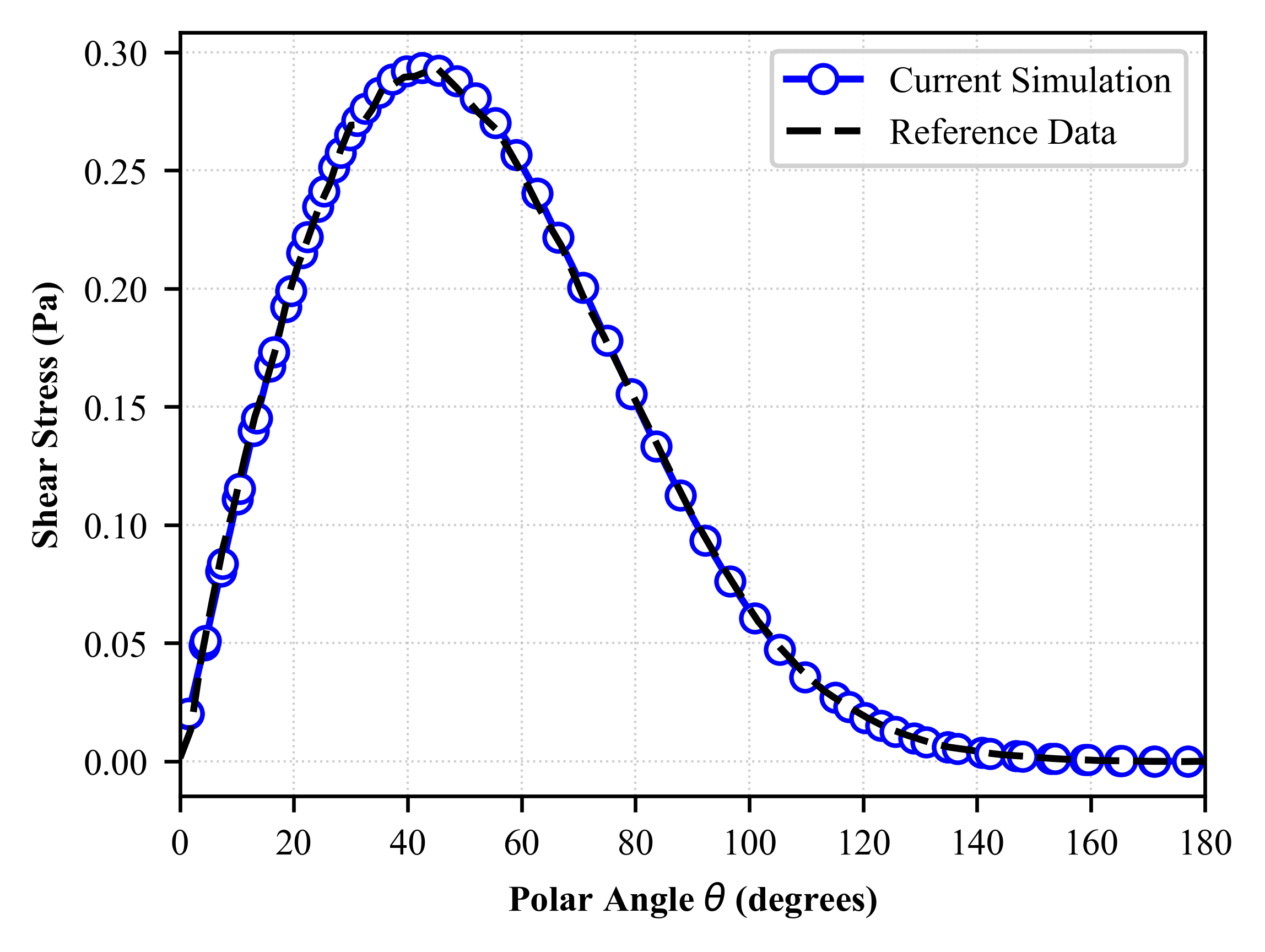}
        \caption{Surface shear stress coefficient}
        \label{fig:EXP2M004-shear_stress_coefficient}
    \end{subfigure}
    \caption{Distributions of surface pressure, heat flux, and shear stress, the reference data is produced by the UGKS code.}
    \label{fig:EXP4M004-surface}
\end{figure}

Figure \ref{fig:Kn0.031-contour} displays the contour distributions of density, velocity, temperature, and local Knudsen number on and around the wall. The local Knudsen number is defined as $Kn_{local} = \frac{\lambda}{\rho/|\nabla\rho|}$, where $\lambda$ is local molecular mean free path. The significant variation in Knudsen number, from 0.031 to values greater than 1 across the entire flow field (Figure \ref{fig:kn0.031-Kn}), highlights the algorithm’s effectiveness in capturing flow physics across different rarefied flow regimes.
The surface quantities are presented in Figure \ref{fig:EXP4M004-surface}. The present results exhibit excellent agreement with the reference data generated by the Unified Gas-Kinetic Scheme (UGKS) code \cite{long2024implicit}, which is well-validated by the Direct Simulation Monte Carlo (DSMC) method. This shows the accuracy of the current solver in the ionization-free condition.

\subsection{Hypersonic Pre-ionized Argon Flow Around a Hemisphere}

In this section, we conduct a three-dimensional simulation of Mach 4.75 argon flow around a hemisphere, replicating the experimental conditions reported by Kranc et al. \cite{cambel1967}. To further verify the code's accuracy, we first compare the results with the experiment's results. Then we change the Knudsen number of argon atoms to see how the control effects vary in different rarefied degrees.

\subsubsection{Validation against experiments}

\begin{figure}[H]
    \centering
    \begin{subfigure}[b]{0.325\textwidth}
    \centering
    \includegraphics[width=1.0\linewidth]{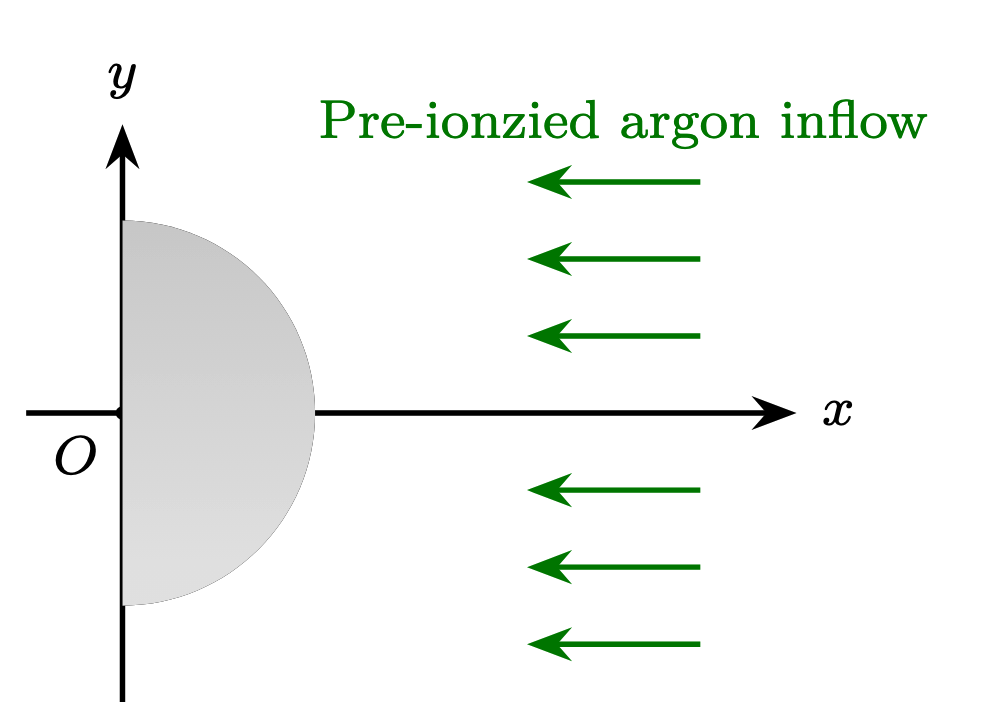}
    \caption{Case setting}
    \label{fig:casesetting}
    \end{subfigure}
    \hfill
    \begin{subfigure}[b]{0.325\textwidth}
    \centering
    \includegraphics[width=0.8\linewidth]{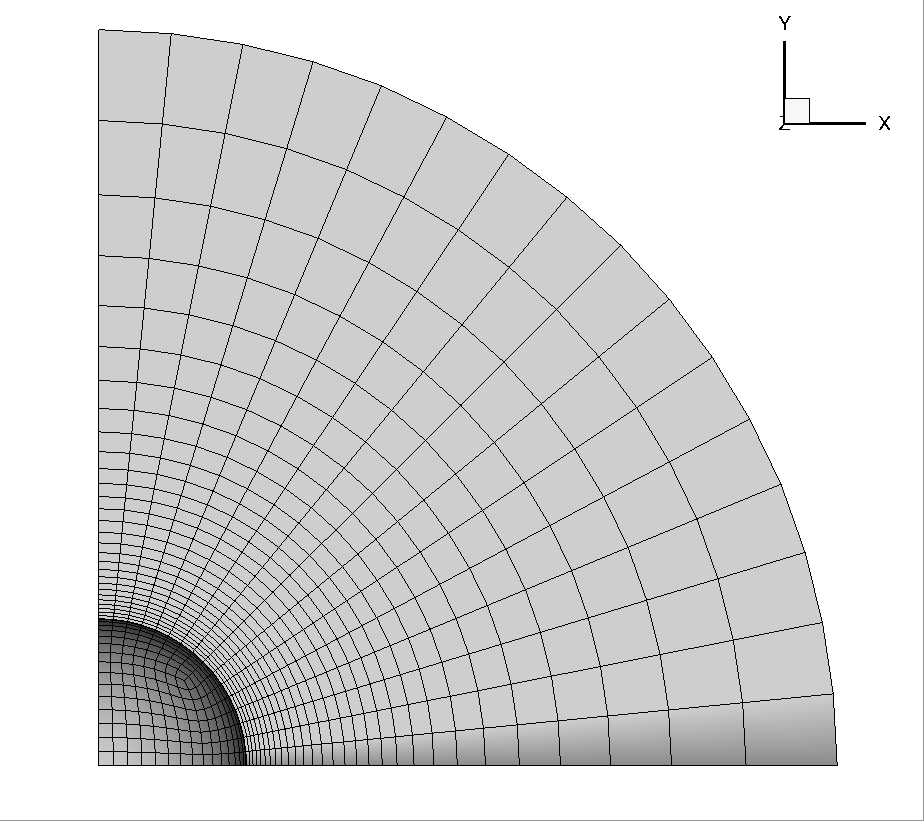}
    \caption{Mesh setting}
    \label{fig:mesh}
    \end{subfigure}
    \hfill 
    \begin{subfigure}[b]{0.325\textwidth}
    \centering
    \includegraphics[width=0.95\linewidth]{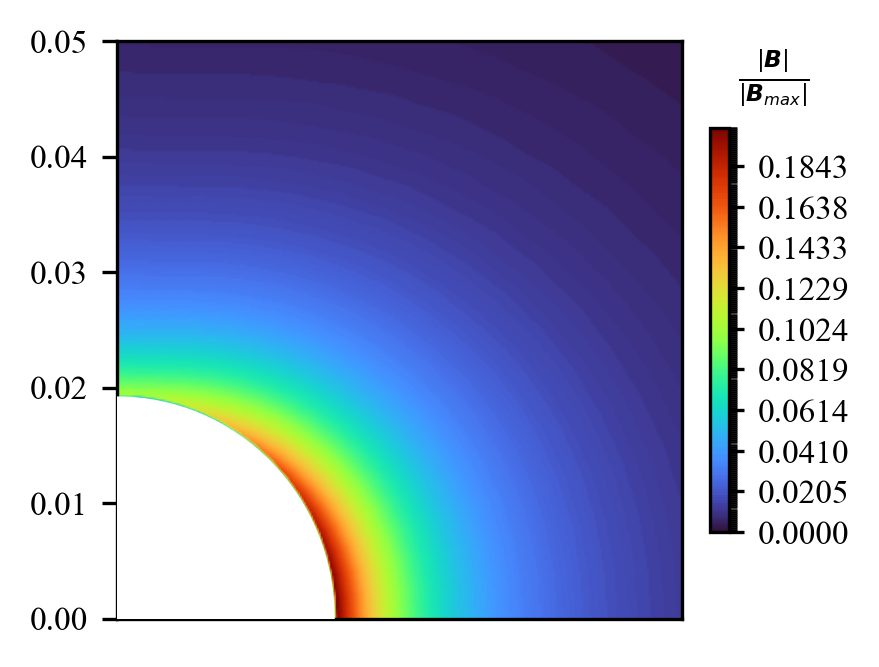}
    \caption{Magnetic field setting $\frac{|\boldsymbol{B}|}{|\boldsymbol{B}_{max}|}$.}
    \label{fig:Bmag-contour}
    \end{subfigure}
    \caption{Computational domain specifications and magnetic magnitude distribution for hypersonic flow over a hemisphere.}
    \label{fig:domain-specification}
\end{figure}

The hemisphere radius is defined as $r_n = 0.01905$ m. The simulation imposes a hypersonic freestream flow in the $-x$ direction incident upon the hemisphere as shown in Figure \ref{fig:casesetting}. $1/4$ of the hemisphere is simulated as shown in Figure \ref{fig:mesh}. The height of the first layer is 0.001m, and the total cell number is 5568.

In the experiments conducted by Kranc et al., the inflow comprised a partially ionized argon flow generated by a plasma torch. Kranc et al. estimated the ionization degree to be $\alpha = 0.025$, and then Bisek et al. \cite{bisek2010numerical}, employing a more refined Saha equation for weakly ionized atomic gases, re-estimated the ionization degree as $\alpha = 0.00623$. This revised value is adopted in the present study. Kranc et al. asserted that the flow remained chemically frozen, allowing the ionization degree to be treated as a constant throughout the flow field. Detailed parameters are listed in Table \ref{tab:hemisphere-parameters}.

\begin{table}[h!]
    \centering
    \begin{tabular}{lc}
        \hline
        \hline
        Parameter & Value \\
        \hline
        Mach number & 4.75 \\
        Knudsen number & 0.044 \\
        Argon atomic mass & $6.63 \times 10^{-26}\ \mathrm{kg}$ \\
        Argon diameter (VHS) & $4.17 \times 10^{-10}\ \mathrm{m}$ \\
        Incoming argon density & $1.09 \times 10^{-4}\ \mathrm{kg/m^3}$ \\
        Incoming argon ion density & $6.85 \times 10^{-7}\ \mathrm{kg/m^3}$ \\
        Incoming electron number density & $1.0332 \times 10^{19}\ \mathrm{m^{-3}}$ \\
        Mass ratio $m_{Ar^+}/m_{e^-}$ & 10 \\
        Characteristic temperature & 1100 K \\
        Incoming electron temperature & 5000 K \\
        Wall temperature & 1100 K \\
        Incoming argon velocity & 3000 m/s \\
        Hemisphere radius &  0.01905 m \\
        \hline
    \end{tabular}
    \caption{Simulation parameters for the rarefied argon flow around a hemisphere.}
    \label{tab:hemisphere-parameters}
\end{table}

A magnetic dipole, with its magnetic moment oriented along the x-axis, is positioned at the center of the hemisphere. Its orientation opposes the incoming flow along the stagnation line. The magnetic field generated by this dipole is defined as:
$$
\mathbf{B}=\frac{B_{\max }}{2\left(\hat{x}^2+\hat{y}^2+\hat{z}^2\right)^{5 / 2}}\left[\begin{array}{c}
2 \hat{x}^2-\left(\hat{y}^2+\hat{z}^2\right) \\
3 \hat{x} \hat{y} \\
3 \hat{x} \hat{z}
\end{array}\right],
$$
where, $\hat{x} = x/r_n$, $\hat{y} = y / r_n$, and $\hat{z} = z / r_n$ represent the coordinates normalized by the hemisphere radius, $r_n$. The parameter $B_{max}$ denotes the maximum magnetic field strength, which is attained at the stagnation point. The distribution of magnetic field magnitude is depicted in Figure \ref{fig:Bmag-contour}. In the following analysis, we employ $B_{max}$ to characterize the magnetic field strength.  Simulations are performed at four distinct magnetic field strengths: 0.223 T, 0.316 T, 0.387 T, and 0.447 T.

In this case, $Ar$-$Ar$, $Ar^{+}$-$Ar^{+}$, $e^{-}$-$e^{-}$ collision is modeled as VHS collision with $\omega=0.81$ and $d=4.17\times {10^{-10}} $m. $Ar$-$Ar^+$, $Ar$-$e^-$, $Ar^+$-$e^-$ cross section is described by fitter curves from Devoto \cite{devoto1973transport} as shown in Table \ref{tab:cross-sections}.
\begin{table}[h!]
    \centering
    \begin{tabular}{lc}
        \hline
        \hline
        Collision pair & Fitted curve \\
        \hline
        $Ar$-$Ar^+$ & $-0.00182T_{Ar^+} + 116.1$ \\
        $Ar$-$e^-$ & $0.0004T_{e^-}0.3838$ \\
        $Ar^+$-$e^-$& $0.001$ \\
        \hline
    \end{tabular}
    \caption{Cross section of Ar-Ar+, Ar-e-, Ar+-e- (unit is $\text{\r{A}}^2$).}
    \label{tab:cross-sections}
\end{table}
Electron mass was artificially increased ($m_{Ar^+}/m_{e^-}=10$) to be computationally tractable. This modification, however, reduces the electrical conductivity. Therefore, the magnetic field magnitude is artifially increased, following the scaling by the similarity parameter $Q=\frac{\sigma^{cond}|\boldsymbol{B}_0|^2L_0}{\rho_0 |\boldsymbol{U}_0|}$, see more explanation in \ref{increaseB}.

\begin{figure}
    \centering
    \begin{subfigure}[b]{0.3\textwidth}
        \centering
        \includegraphics[width=0.9\textwidth]{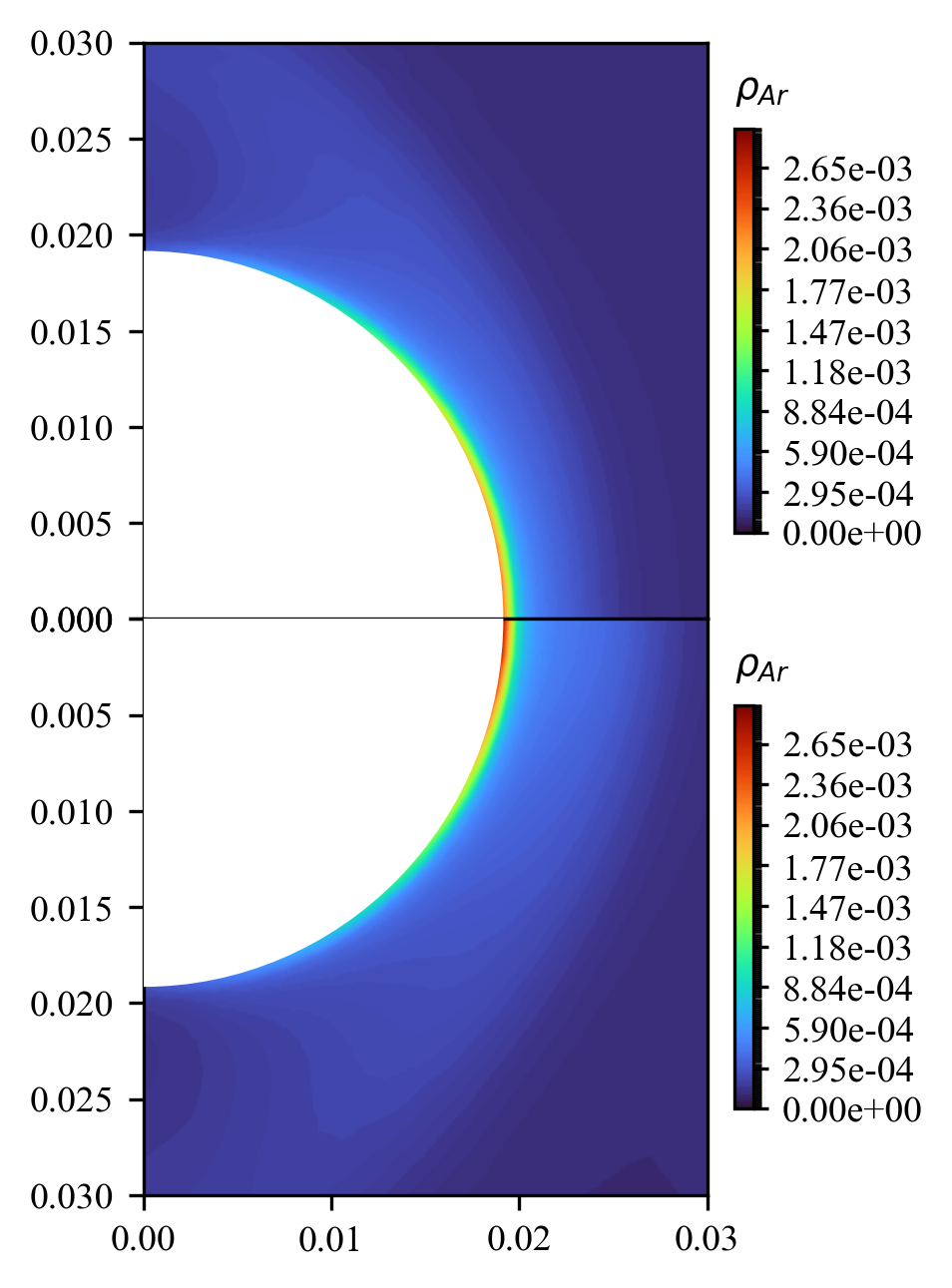}
        \caption{}
        \label{fig:rhoAr}
    \end{subfigure}
    \hfill 
    \begin{subfigure}[b]{0.3\textwidth}
        \centering
        \includegraphics[width=0.9\textwidth]{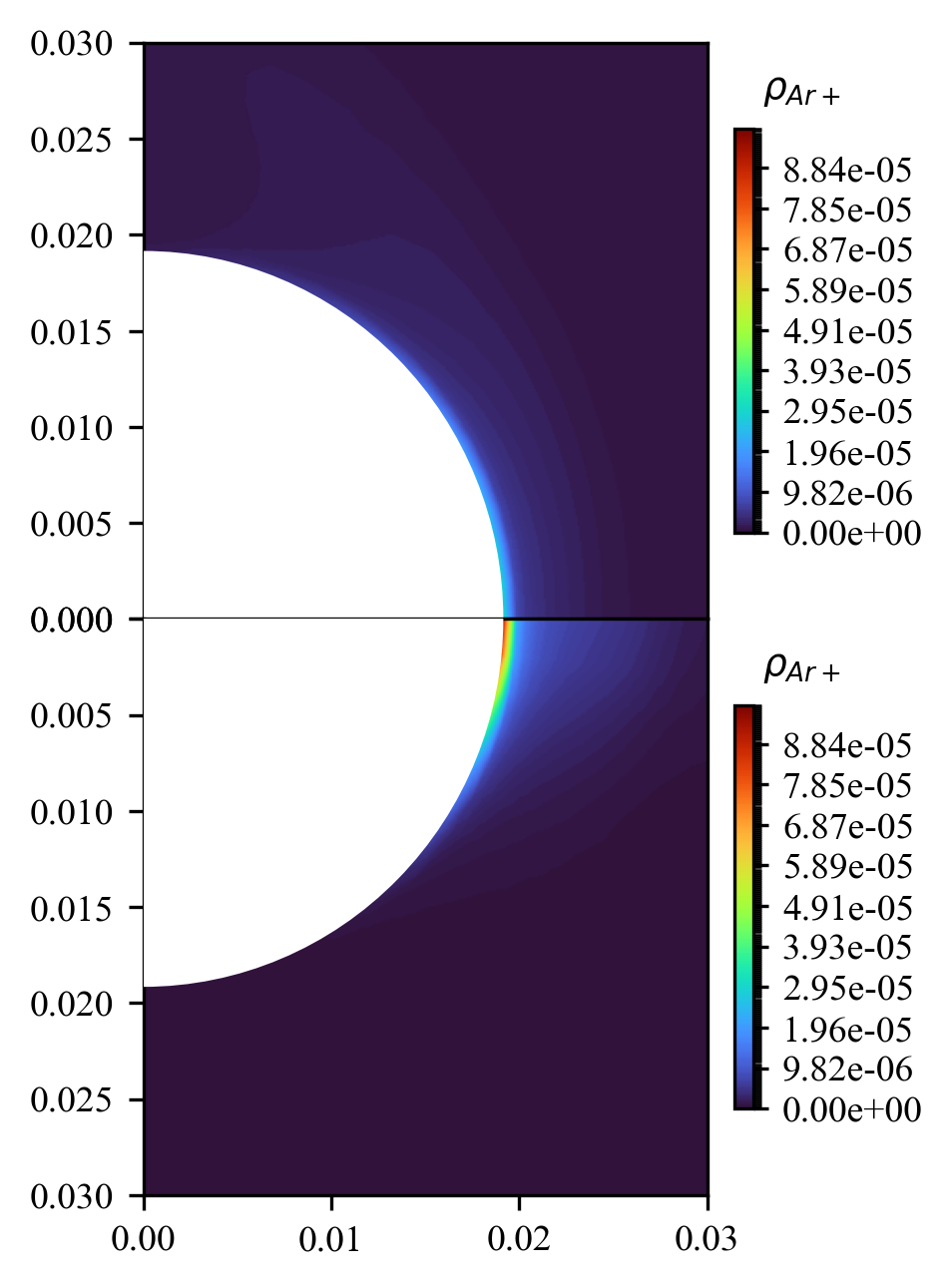}
        \caption{}
        \label{fig:rhoArIon}
    \end{subfigure}
    \hfill 
    \begin{subfigure}[b]{0.3\textwidth}
        \centering
        \includegraphics[width=0.9\textwidth]{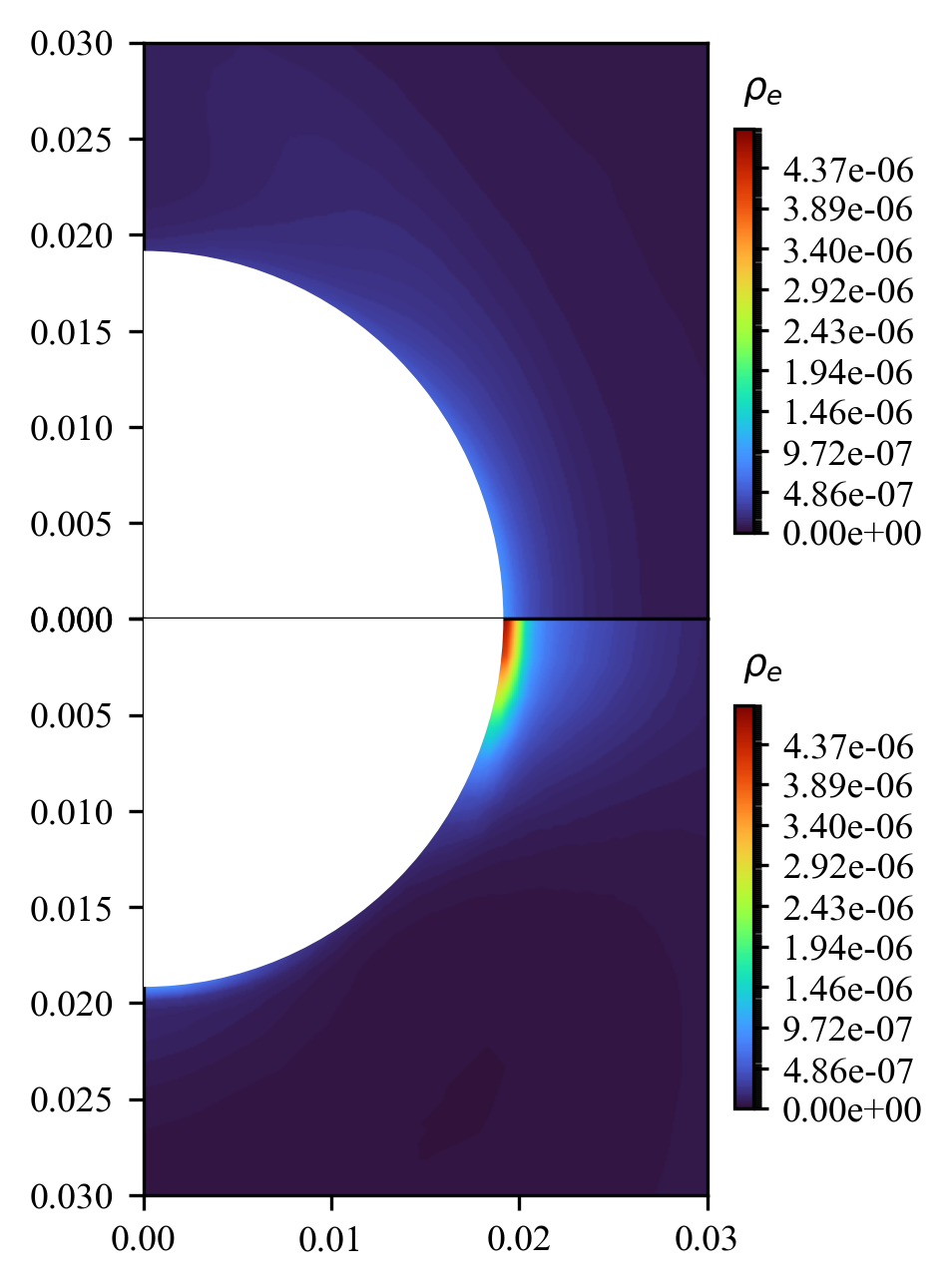}
        \caption{}
        \label{fig:rhoelectron}
    \end{subfigure}
    \vfill
    \begin{subfigure}[b]{0.3\textwidth}
        \centering
        \includegraphics[width=0.9\textwidth]{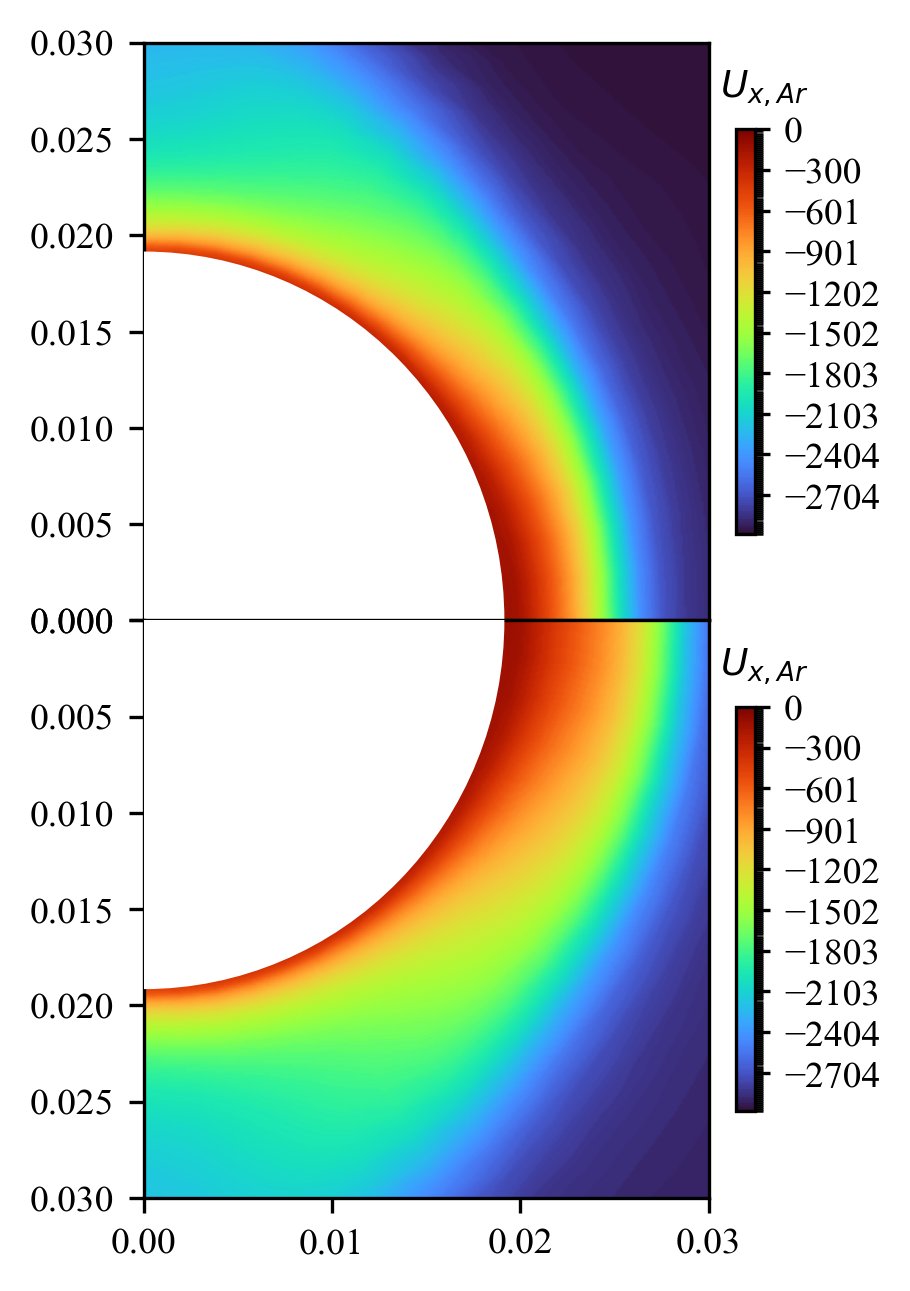}
        \caption{}
        \label{fig:UxAr}
    \end{subfigure}
    \hfill 
    \begin{subfigure}[b]{0.3\textwidth}
        \centering
        \includegraphics[width=0.9\textwidth]{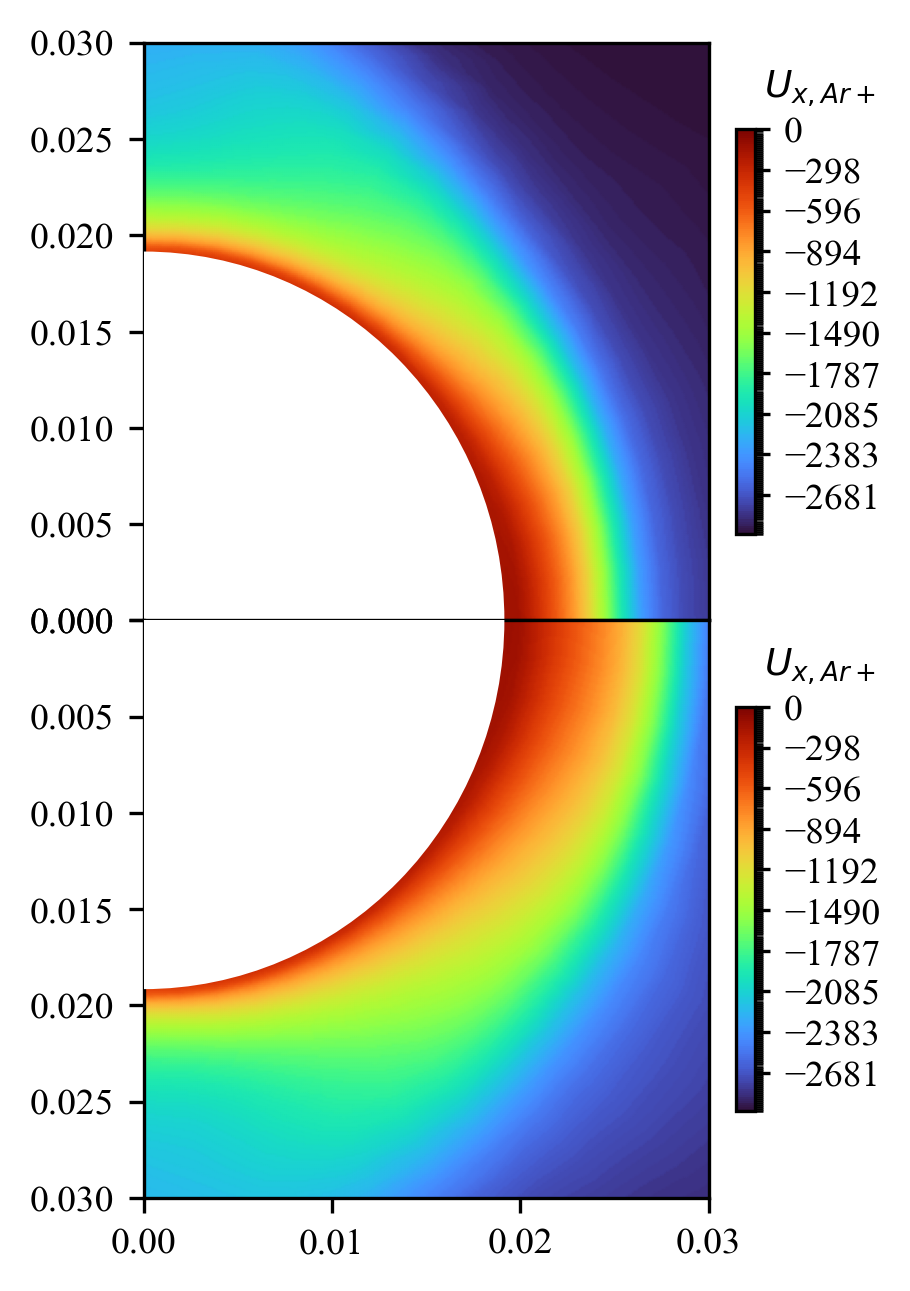}
        \caption{}
        \label{fig:UxArIon}
    \end{subfigure}
    \hfill 
    \begin{subfigure}[b]{0.3\textwidth}
        \centering
        \includegraphics[width=0.9\textwidth]{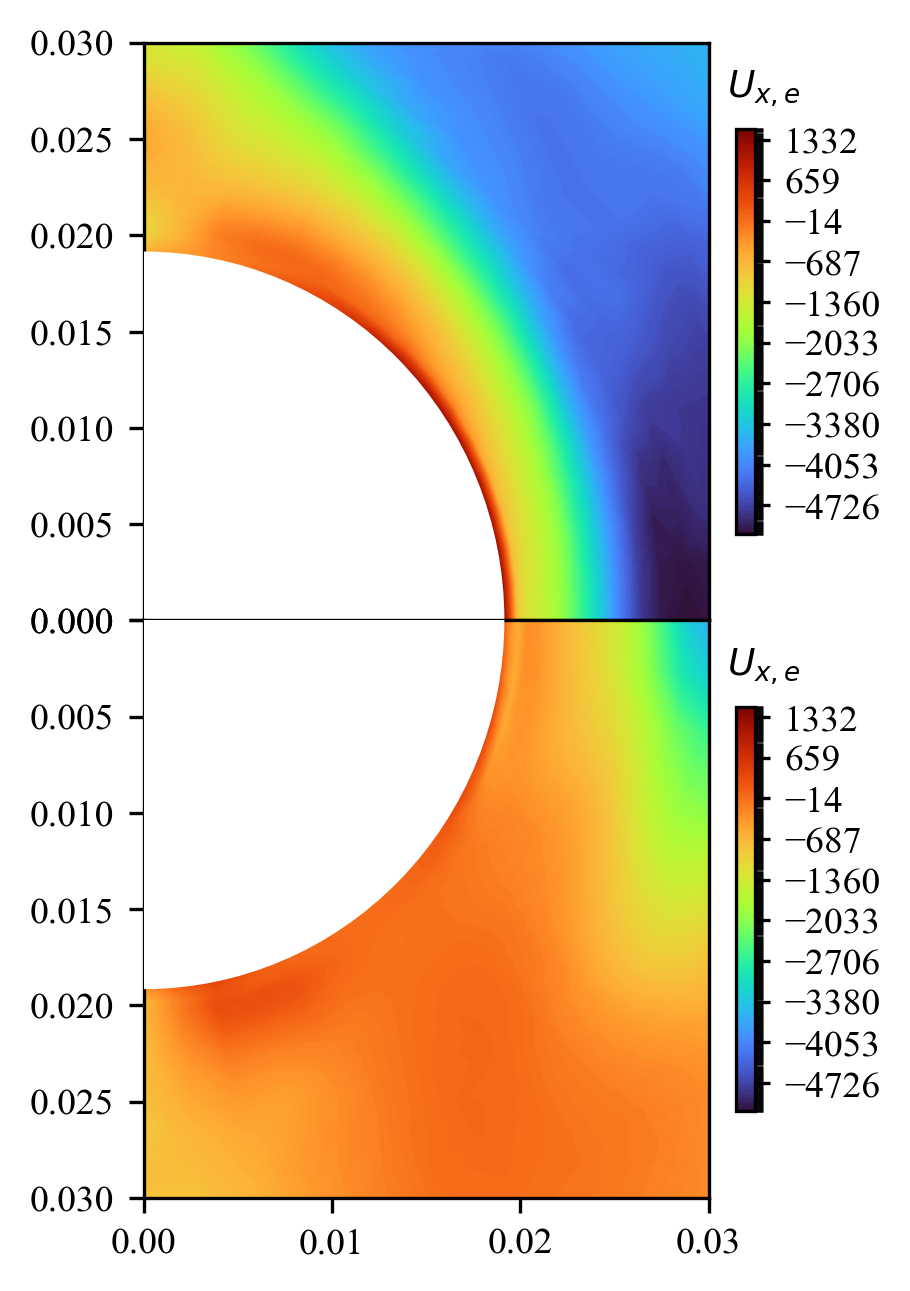}
        \caption{}
        \label{fig:Uxelectron}
    \end{subfigure}
    \vfill
    \begin{subfigure}[b]{0.3\textwidth}
        \centering
        \includegraphics[width=0.9\textwidth]{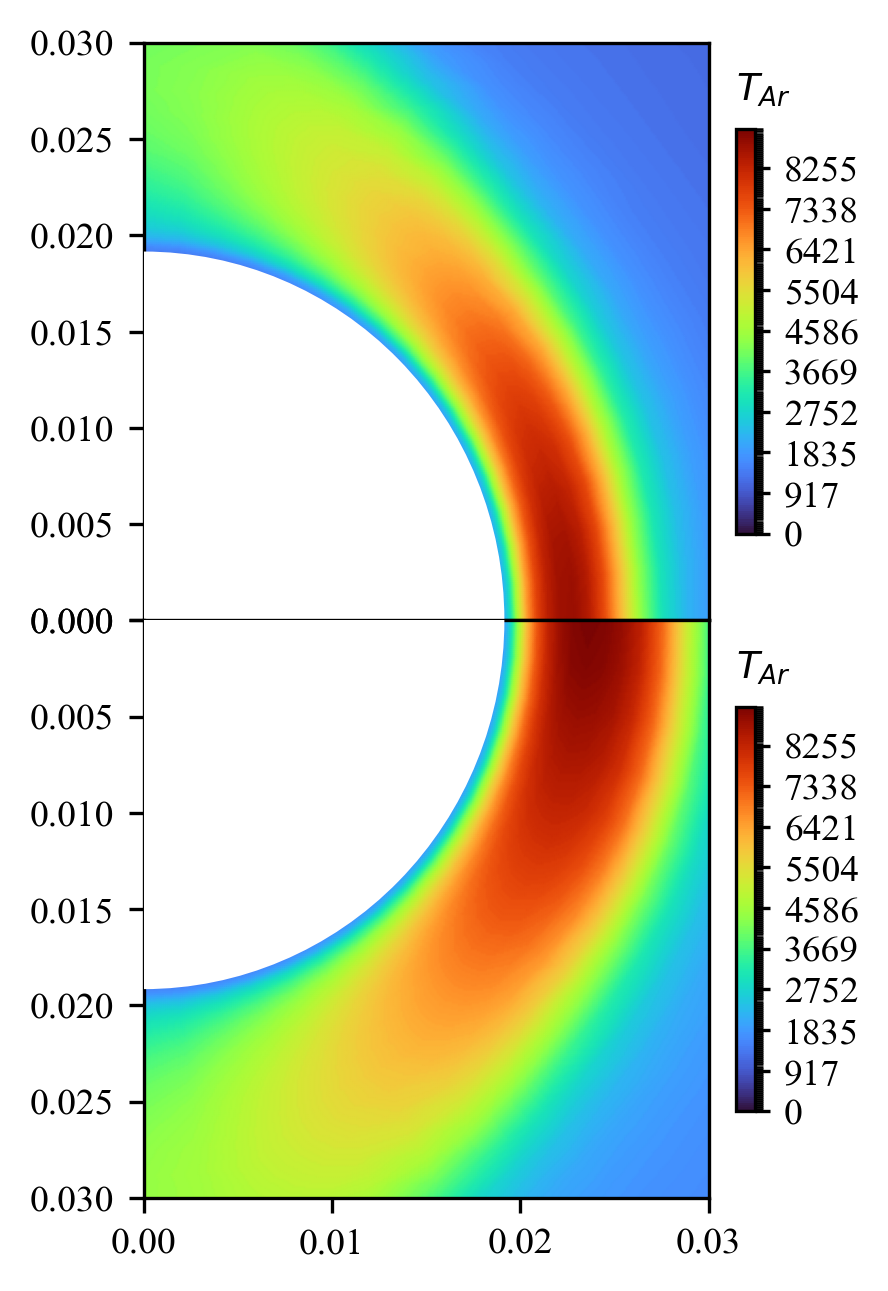}
        \caption{}
        \label{fig:TAr}
    \end{subfigure}
    \hfill 
    \begin{subfigure}[b]{0.3\textwidth}
        \centering
        \includegraphics[width=0.9\textwidth]{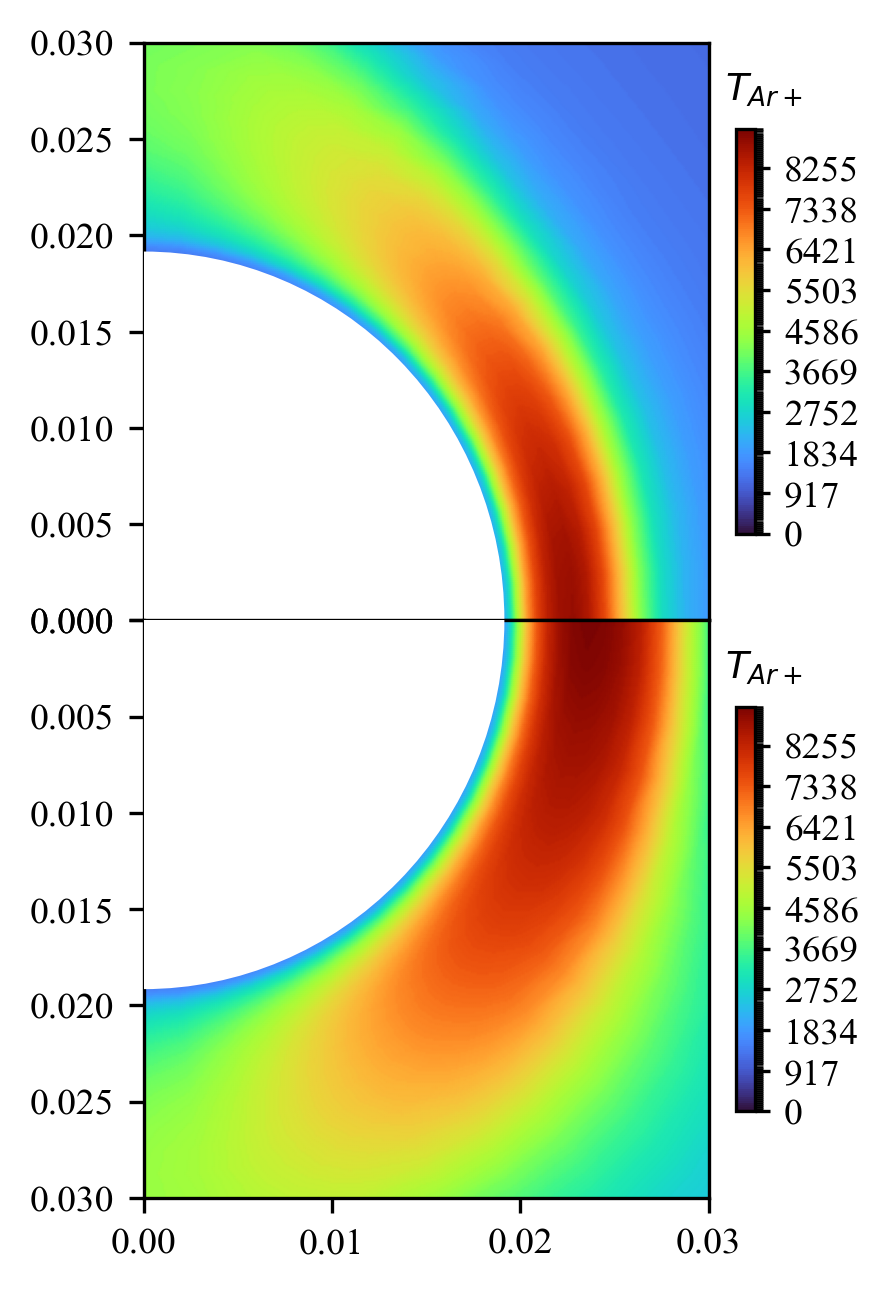}
        \caption{}
        \label{fig:TArIon}
    \end{subfigure}
    \hfill 
    \begin{subfigure}[b]{0.3\textwidth}
        \centering
        \includegraphics[width=0.9\textwidth]{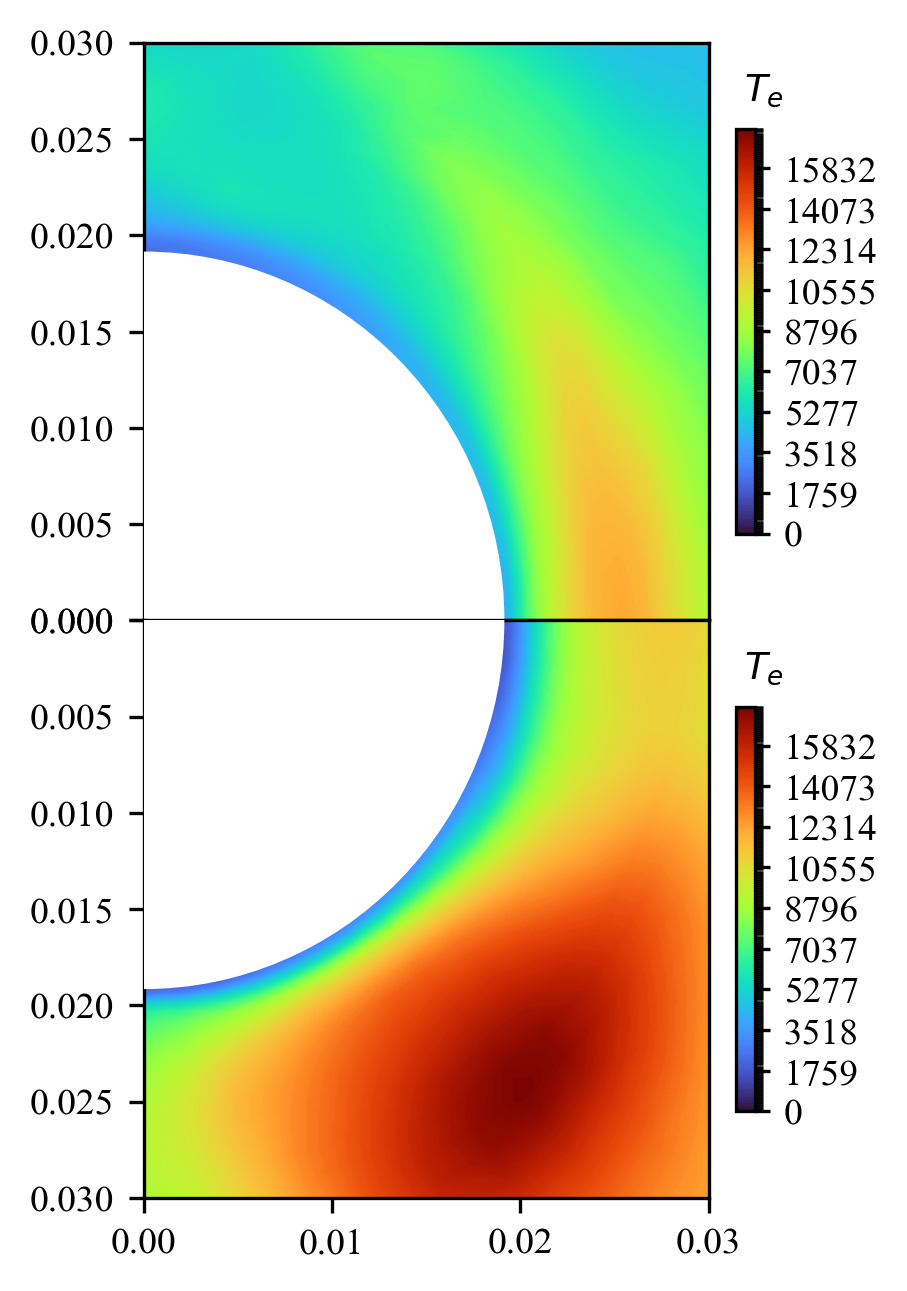}
        \caption{}
        \label{fig:Telectron}
    \end{subfigure}
    \caption{Contour plots of mass density without external magnetic field (upper) and with external magnetic field (lower) in the figures, where (a) $Ar$ (b) $Ar^+$ (c) $e^-$ and  Knudsen number of $Ar$ and x-velocity of (d) $Ar$ (e) $Ar^+$ (f) $e^-$ and temperature of (g) $Ar$ (h) $Ar^+$ (i) $e^-$ at $B_{max}=0.447T$.}
    \label{fig:EXP3EXP014-compare}
\end{figure}

\begin{figure}
    \centering
    \begin{subfigure}[b]{0.3\textwidth}
        \centering
        \includegraphics[width=0.9\textwidth]{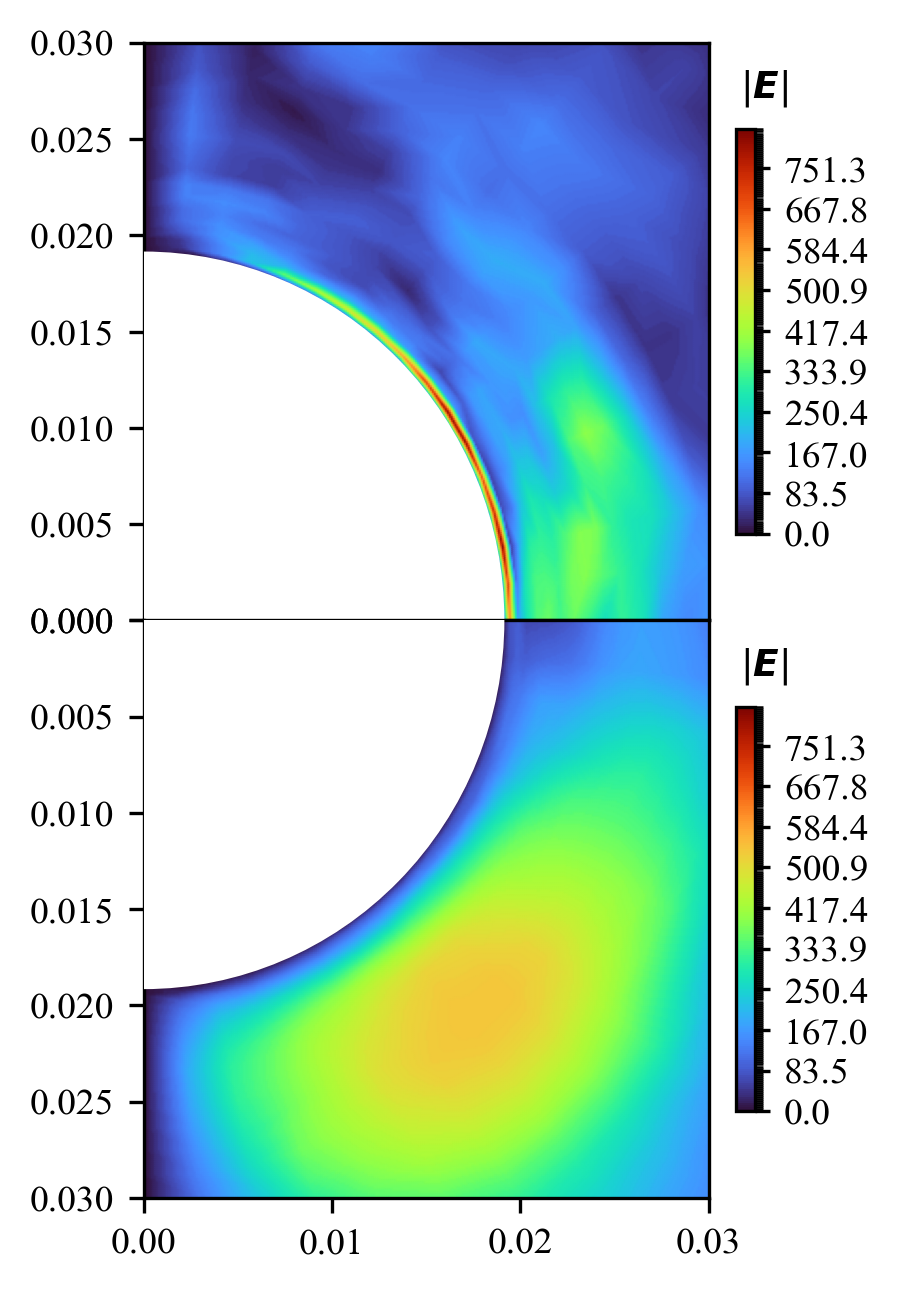}
        \caption{}
        \label{fig:Emag}
    \end{subfigure}
    \hfill 
    \begin{subfigure}[b]{0.3\textwidth}
        \centering
        \includegraphics[width=0.9\textwidth]{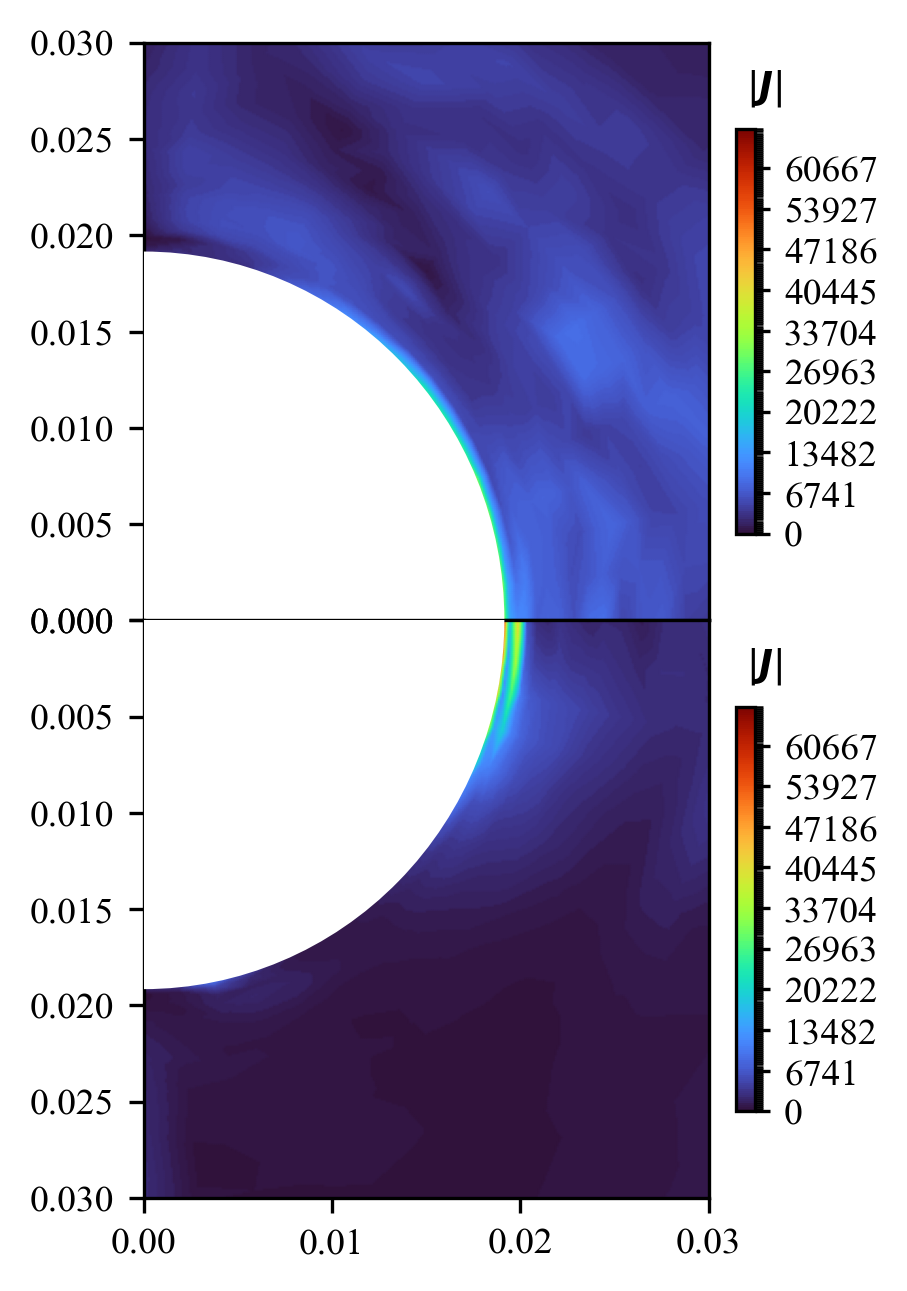}
        \caption{}
        \label{fig:Jmag}
    \end{subfigure}
    \hfill 
    \begin{subfigure}[b]{0.3\textwidth}
        \centering
        \includegraphics[width=0.9\textwidth]{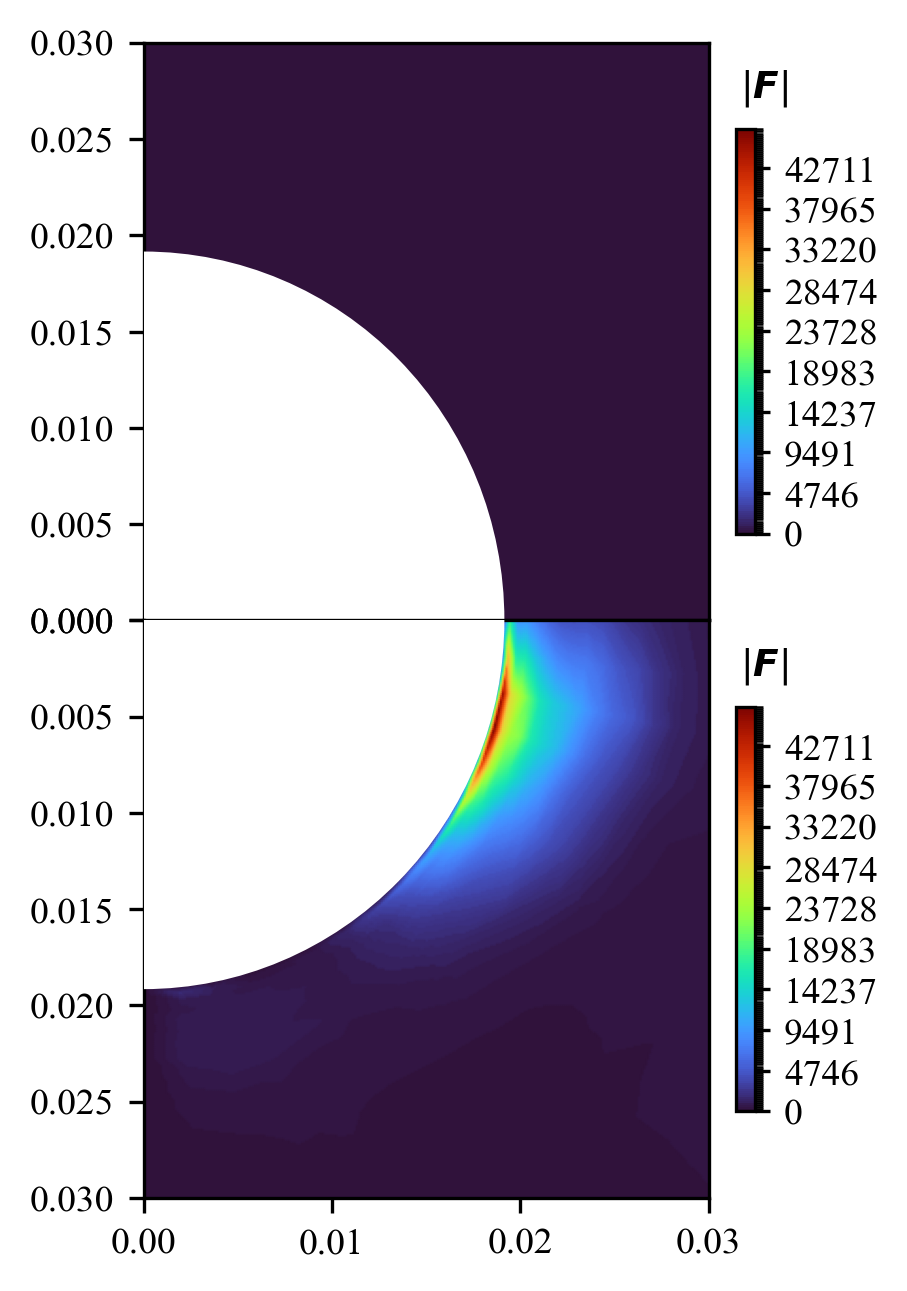}
        \caption{}
        \label{fig:Fmag}
    \end{subfigure}
    \caption{Contour plots of (a) electric field magnitude $|\boldsymbol{E}|$ (b) current density magnitude $\boldsymbol{J}$ (c)  Lorentz force magnitude $|\boldsymbol{F}|$ at $B_{max}=0.447T$. Upper and lower figures refer to the results without and with external magnetic fields.}
    \label{fig:EXP3EXP014-compare2}
\end{figure}

Current formulation allows for the observation of distinct behaviors among atoms, ions, and electrons, including velocity and temperature slip arising from rarefied effects.
Figures \ref{fig:EXP3EXP014-compare} and \ref{fig:EXP3EXP014-compare2} display contour distributions, with upper halves showing simulations without an external magnetic field and lower halves depicting results with background magnetic field strengths ($B=0.447$ T). The data confirm that an applied magnetic field enhances shock standoff distance. Figures \ref{fig:rhoArIon} and \ref{fig:rhoelectron} demonstrate that after the application of a magnetic field, both ions and electrons are significantly trapped near the stagnation point. This trapping occurs due to the localized maximum strong magnetic field at this location. The collision coupling between these charged particles and argon atoms also contributes to a slight accumulation of argon atoms in the stagnation region, as depicted in Figure \ref{fig:rhoAr}. Figures \ref{fig:UxAr}, \ref{fig:UxArIon}, and \ref{fig:Uxelectron} present the x-velocity of $Ar$, $Ar^+$, and $e^-$. In the stagnation region, $Ar^+$ velocity closely follows $Ar$ velocity. In the shoulder region \footnote{The shoulder region comprises arc degrees spanning from approximately $30^\circ$  to $60^\circ$, with the stagnation line serving as the $0^\circ$ reference.}, outside of shock, their velocities diverge due to the magnetic field influence on $Ar^+$. Electron velocity deviates significantly from heavy particles as the magnetic field more easily accelerates them, and the velocity decoupling is further promoted by low collision frequencies in rarefied environments. Figures \ref{fig:TAr} to \ref{fig:Telectron} show similar decoupling in temperature.

The present formulation permits the observation of electric fields generated by diverse mechanisms, contrasting with MHD-based methods that primarily consider resistivity.
Figures \ref{fig:Emag} depict the electric field magnitude with and without an external magnetic field. Without a magnetic field, the primary electric field is observed in the stagnation region, driven by a strong pressure gradient across the shock front that separates electrons and ions. The non-neutral region is amplified here by an increased Debye length. In reality, this non-neutrality is localized to the shock front and wall sheath. The application of a magnetic field induces an additional charge separation in the shoulder region due to the different responses of ions and electrons to the magnetic field.  Figure \ref{fig:Jmag} shows a strong current density concentrated in the stagnation region. The current density here is computed by $\boldsymbol{J} = q_in_i\boldsymbol{U}_i + q_en_e\boldsymbol{U}_e$.
 Figure \ref{fig:Fmag} reveals a substantial Lorentz force near this area. The Lorentz force is computed as $\boldsymbol{F} = \boldsymbol{J}\times \boldsymbol{B}$. The force does not precisely align with the peak current density and magnetic field strength because the magnitude of the Lorentz force $\boldsymbol{J}\times\boldsymbol{B}$ is also critically dependent on the angle between the current density $\boldsymbol{J}$ and the magnetic field $\boldsymbol{B}$.

\begin{figure}[H]
    \centering
    \begin{subfigure}[b]{0.58\textwidth}
        \centering
        \includegraphics[width=0.9\linewidth]{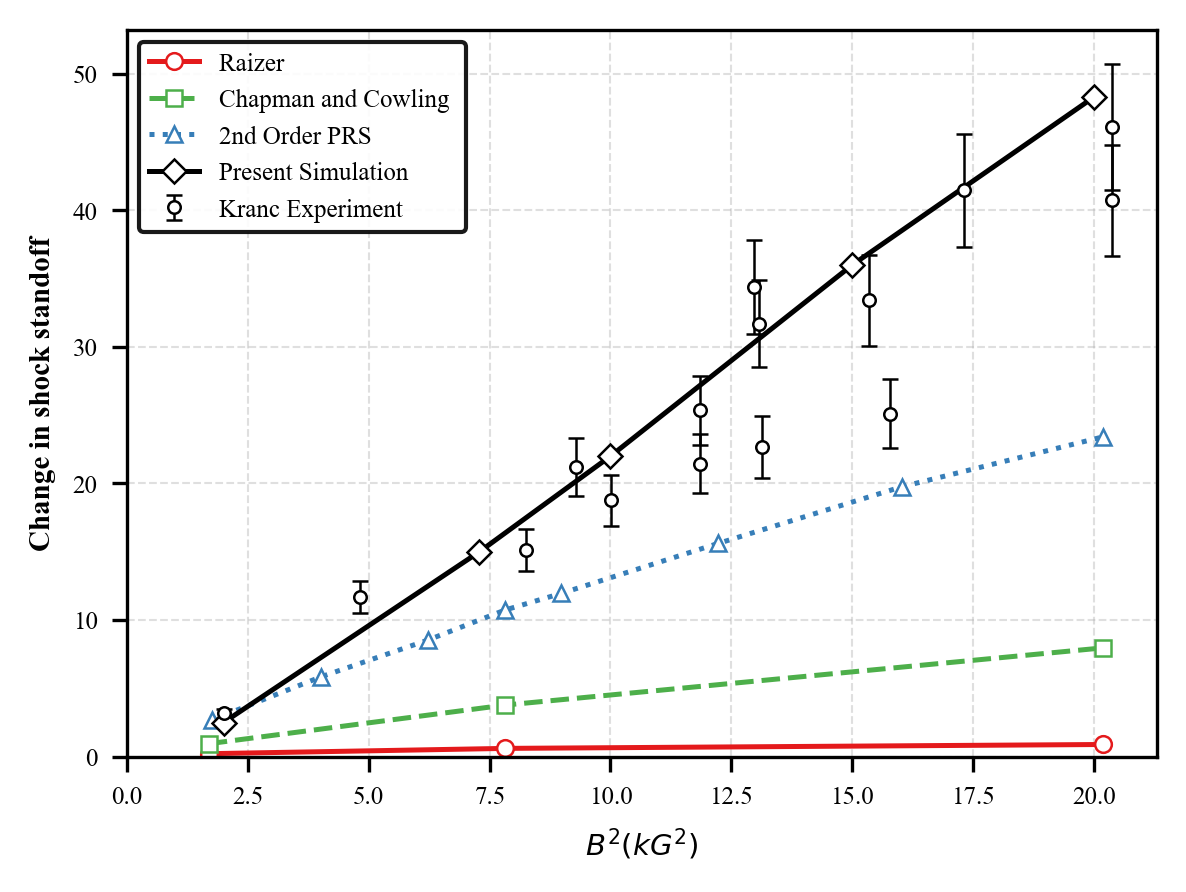}
        \caption{}
    \end{subfigure}
    \caption{ Changes in shock standoff distance as a function of magnetic field strength for Mach 4.75 pre-ionized argon flow around a hemispherical geometry. \cite{cambel1967} (Experimental uncertainties is $\pm 10\%$).}
    \label{fig:shock-distance}
\end{figure}

The dependence of shock standoff distance on applied magnetic field strength is presented in Figure \ref{fig:shock-distance}. The shock front location is determined using the density jump ratio, which is defined as \cite{bisek2010numerical} :
$$
\lim_{M_1\rightarrow \infty} \frac{\rho_2}{\rho_1} = \frac{\gamma + 1}{\gamma - 1}
$$
Here, $M_1$ denotes the upstream Mach number, $\gamma$ represents the ratio of specific heats, $\rho_1$ is the upstream density, and $\rho_2$ is the downstream density. For argon gas, with $\gamma_{Ar} = 5/3$, this density ratio is calculated to be 4. Our simulations exhibit good agreement with experimental data. In addition to experimental results, reference simulation data from Bisek et al. \cite{bisek2010numerical}, employing three distinct conductivity models, are included for comparison. The primary distinction between the present formulation and Bisek's work lies in our consideration of non-equilibrium transport phenomena in ionized gases, whereas Bisek's results are based on the MHD equations. For hypersonic flow around a hemisphere, the standoff distance increases with increasing Knudsen number when rarefied effects are significant \cite{long2024nonequilibrium}.

\begin{figure}[H]
    \centering
    \begin{subfigure}[b]{0.48\textwidth}
        \centering
        \includegraphics[width=0.9\linewidth]{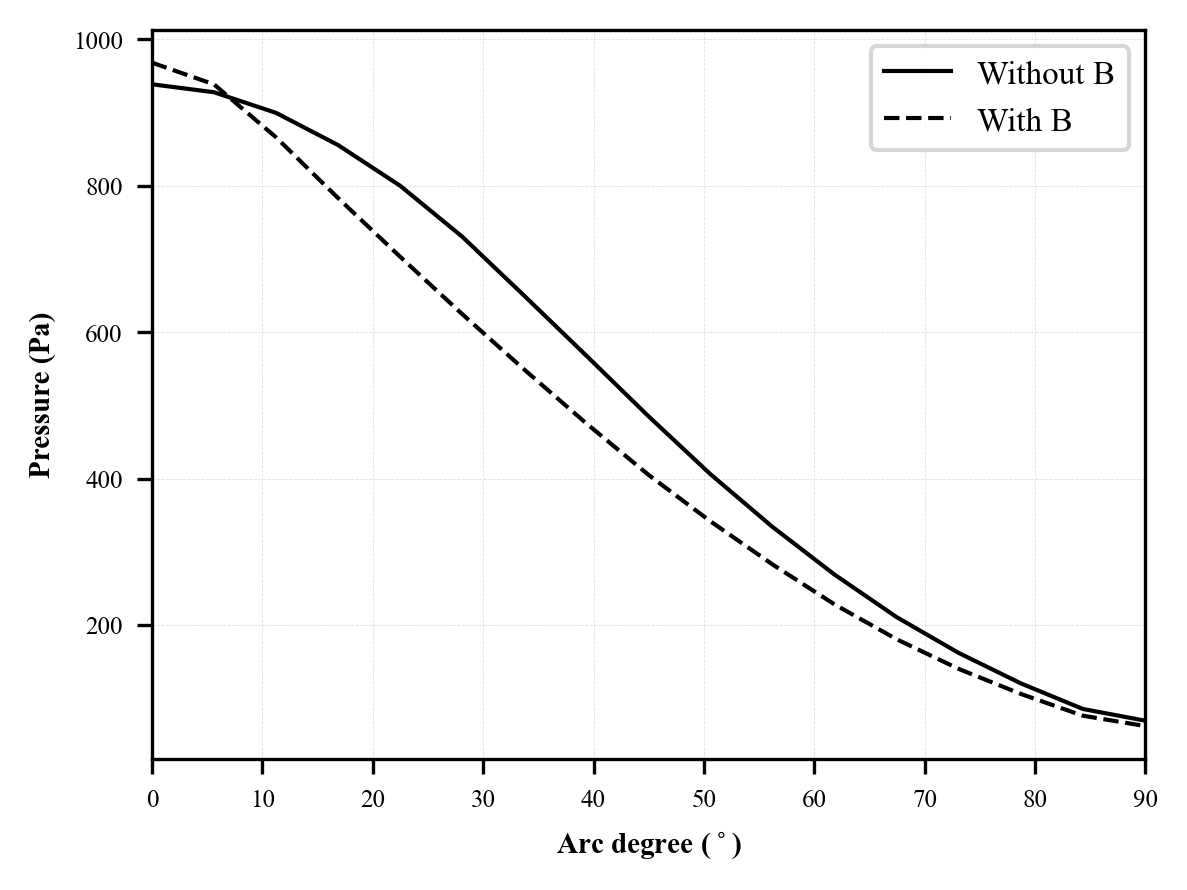}
        \caption{}
        \label{fig:PAr}
    \end{subfigure}
    \hfill
    \begin{subfigure}[b]{0.48\textwidth}
        \centering
        \includegraphics[width=0.9\linewidth]{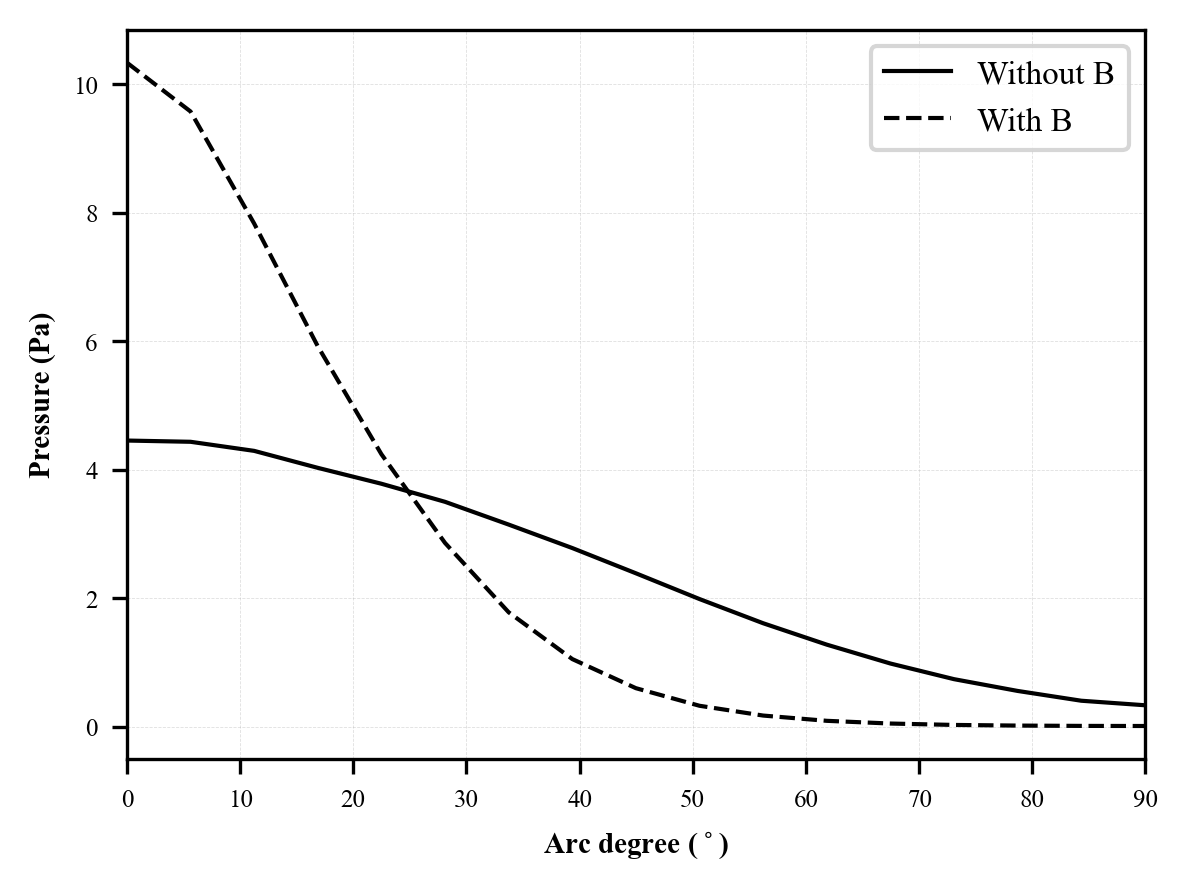}
        \caption{}
        \label{fig:PArIon}
    \end{subfigure}
    \vfill
    \begin{subfigure}[b]{0.48\textwidth}
        \centering
        \includegraphics[width=0.9\linewidth]{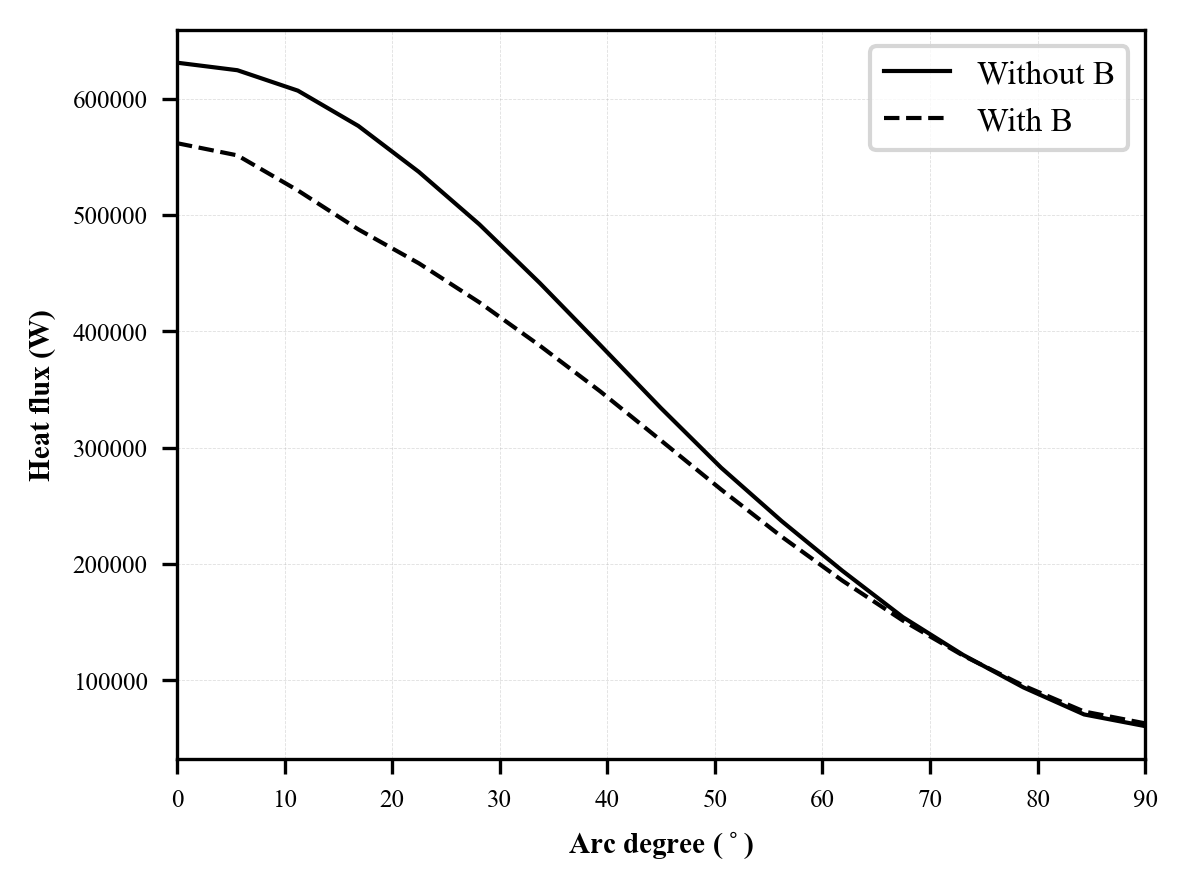}
        \caption{}
        \label{fig:qAr}
    \end{subfigure}
    \hfill
    \begin{subfigure}[b]{0.48\textwidth}
        \centering
        \includegraphics[width=0.9\linewidth]{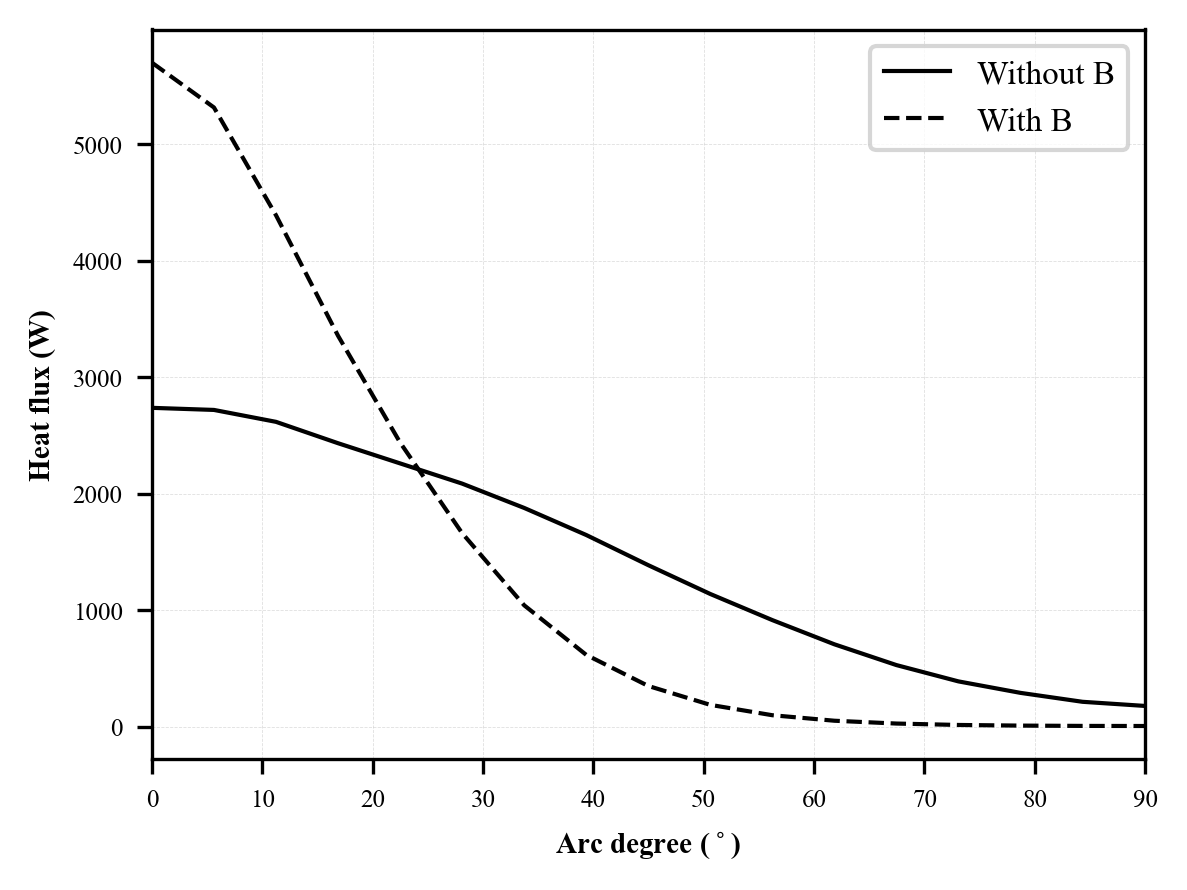}
        \caption{}
        \label{fig:qArIon}
    \end{subfigure}
    \caption{Surface pressure of (a) $Ar$ (b) $Ar^+$ and surface heat flux of (c) $Ar$ (d) $Ar^+$ at $B_{max}=0.447T$.}
    \label{fig:surface}
\end{figure}

Figure \ref{fig:surface} presents the surface pressure and heat flux distributions for both $Ar$ and $Ar^+$. Figure \ref{fig:PAr} illustrates that the surface pressure of $Ar$ increases near the stagnation region and decreases in the shoulder region. A similar trend in surface pressure is observed for $Ar^+$. Figure \ref{fig:qAr} shows a decrease in heat flux near the stagnation region for $Ar$, attributable to an enlarged standoff distance leading to a smaller temperature gradient. In contrast, the heat flux of $Ar^+$ increases in the stagnation region, potentially due to larger kinetic energy caused by the magnetic field. Given that the mole fraction of $Ar^+$ is only 0.00623, the overall surface pressure and heat flux are predominantly governed by $Ar$. The observed increase in stagnation pressure and decrease in heat flux on the hemisphere's surface are consistent with most established literature in this field  \cite{bush1958, poggie2002, fujino2008}.

\subsubsection{Control effects under different Knudsen numbers}

This subsection investigates the influence of control parameters across varying Knudsen numbers (0.2, 0.044, and 0.008). The Knudsen number is determined by the incoming argon gas density, which is set to 2.18$\times \text{10}^{-5}$ kg$\cdot$m$^{-3}$ for Kn = 0.2, 1.09$\times \text{10}^{-4}$ kg$\cdot$m$^{-3}$ for Kn = 0.044, and 5.45$\times \text{10}^{-4}$ kg$\cdot$m$^{-3}$ for Kn = 0.008. The computational geometry, meshing, and other detailed parameters remain the same as in the prior sections. The similarity parameter, $Q$, is held constant across these variations in Knudsen number.  According to Eq. \eqref{eq:conductivity_experssion} and \eqref{eq:interaction_parameter} shown in Eq. \eqref{eq:conductivity_experssion}, if the argon atom density increases by a factor $a$, the parameter $Q$ inherently decreases by a factor $a^2$. To counteract this effect and maintain a constant $Q$, the magnetic field magnitude $|\boldsymbol{B}|$ must be amplified by the same factor $a^2$. The magnetic field magnitude in this section is set to be $B_{max}=0.447T$.

\begin{figure}
    \centering
    \begin{subfigure}[b]{0.3\textwidth}
        \centering
        \includegraphics[width=0.9\textwidth]{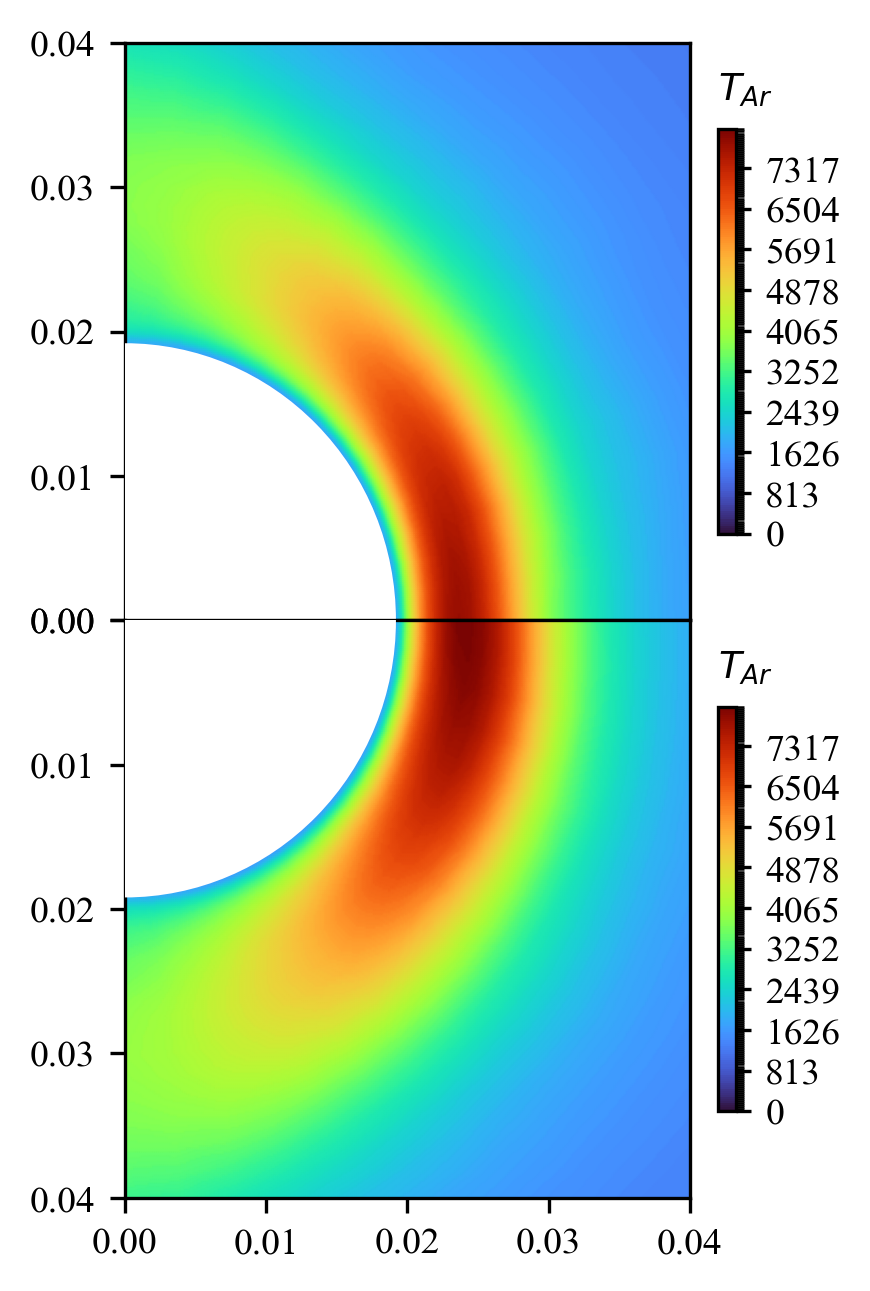}
        \caption{}
        \label{fig:Kn0.2TAr}
    \end{subfigure}
    \hfill 
    \begin{subfigure}[b]{0.3\textwidth}
        \centering
        \includegraphics[width=0.9\textwidth]{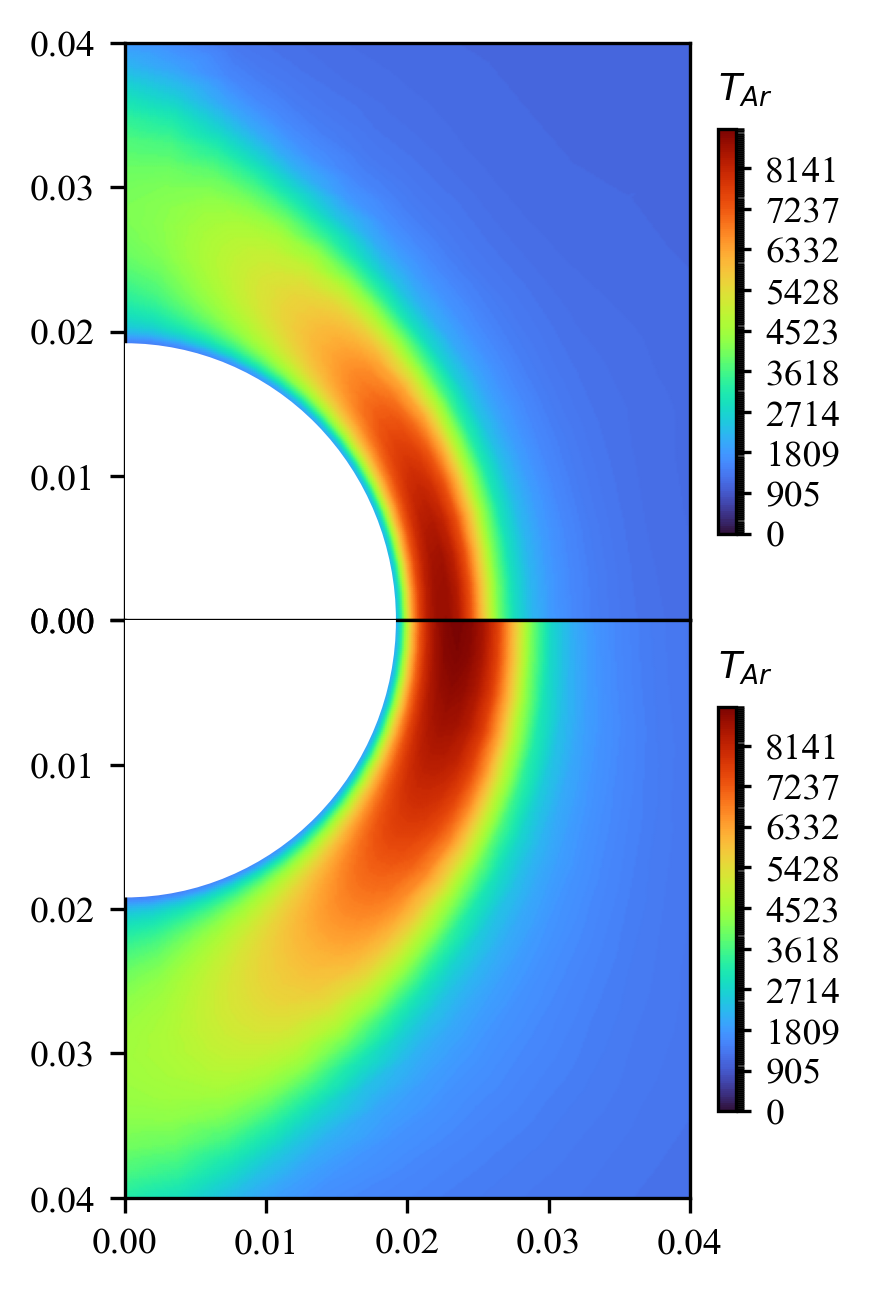}
        \caption{}
        \label{fig:Kn0.044TAr}
    \end{subfigure}
    \hfill 
    \begin{subfigure}[b]{0.3\textwidth}
        \centering
        \includegraphics[width=0.9\textwidth]{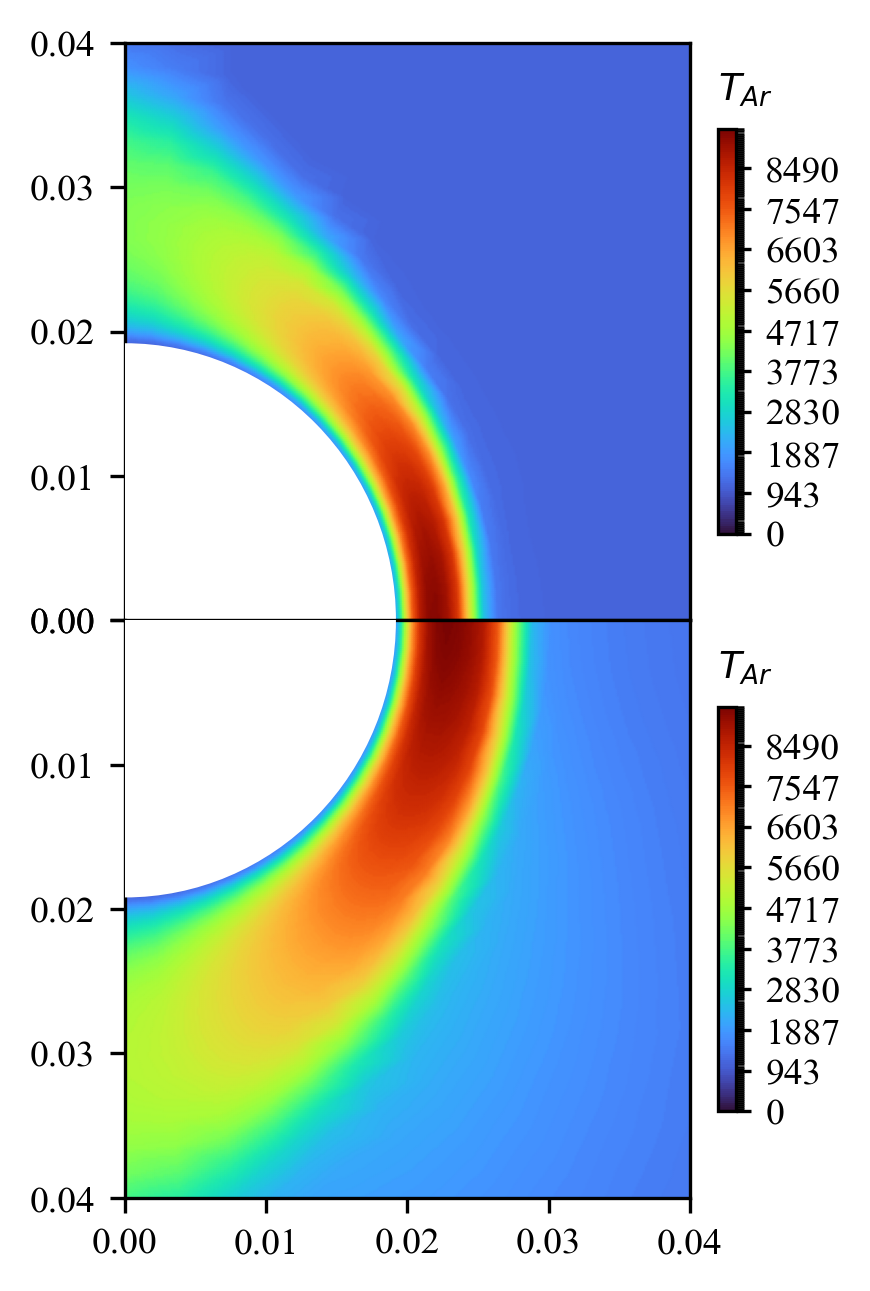}
        \caption{}
        \label{fig:Kn0.008TAr}
    \end{subfigure}
    \caption{Contour plots of the argon atoms temperature field of (a) Kn=0.2 (b) Kn=0.044 (c) Kn=0.008 at $B_{max}=0.447T$. 
    Upper and lower figures refer to the results without and with external magnetic fields.}
    \label{fig:different_Kn_contour}
\end{figure}

\begin{figure}
    \centering
    \begin{subfigure}[b]{0.48\textwidth}
    \centering
        \includegraphics[width=\textwidth]{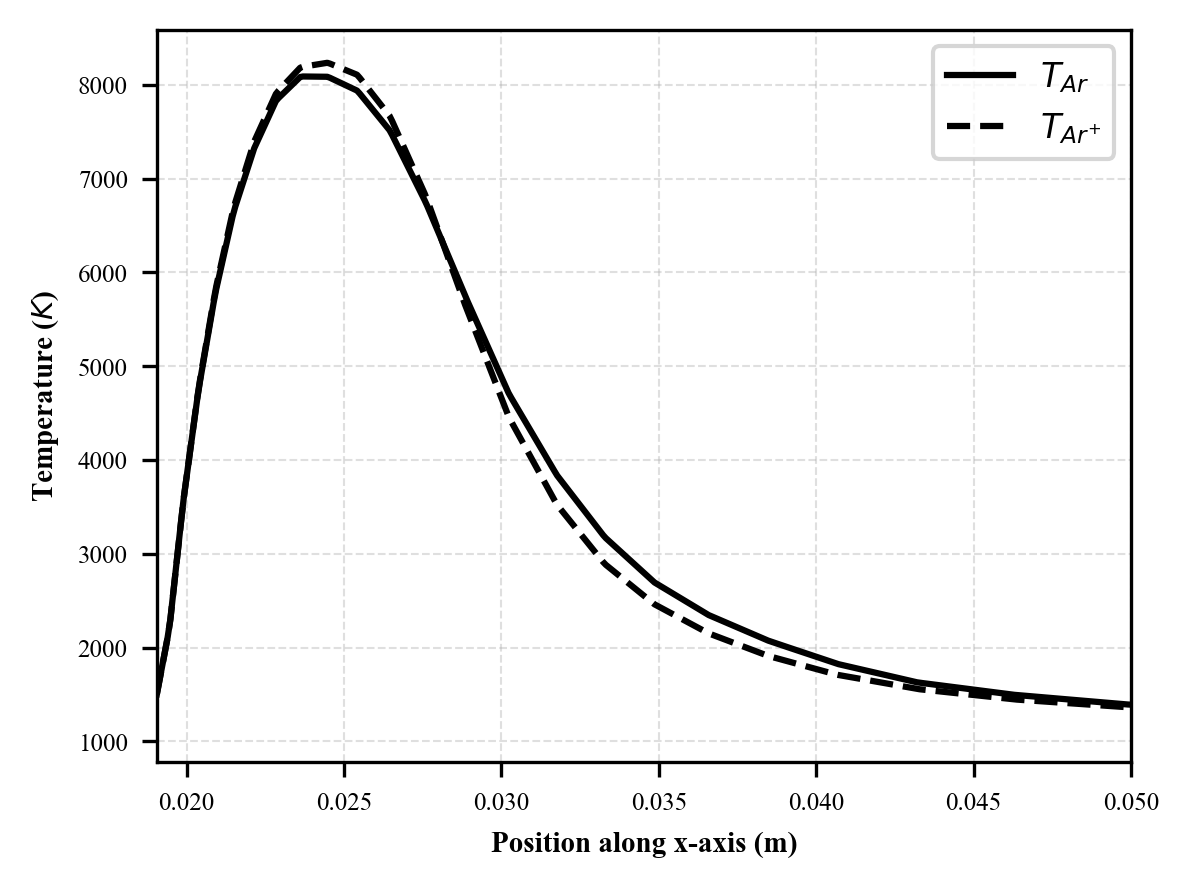}
        \caption{}
        \label{fig:Kn0.2T}
    \end{subfigure}
    \hfill 
    \begin{subfigure}[b]{0.48\textwidth}
    \centering
        \includegraphics[width=\textwidth]{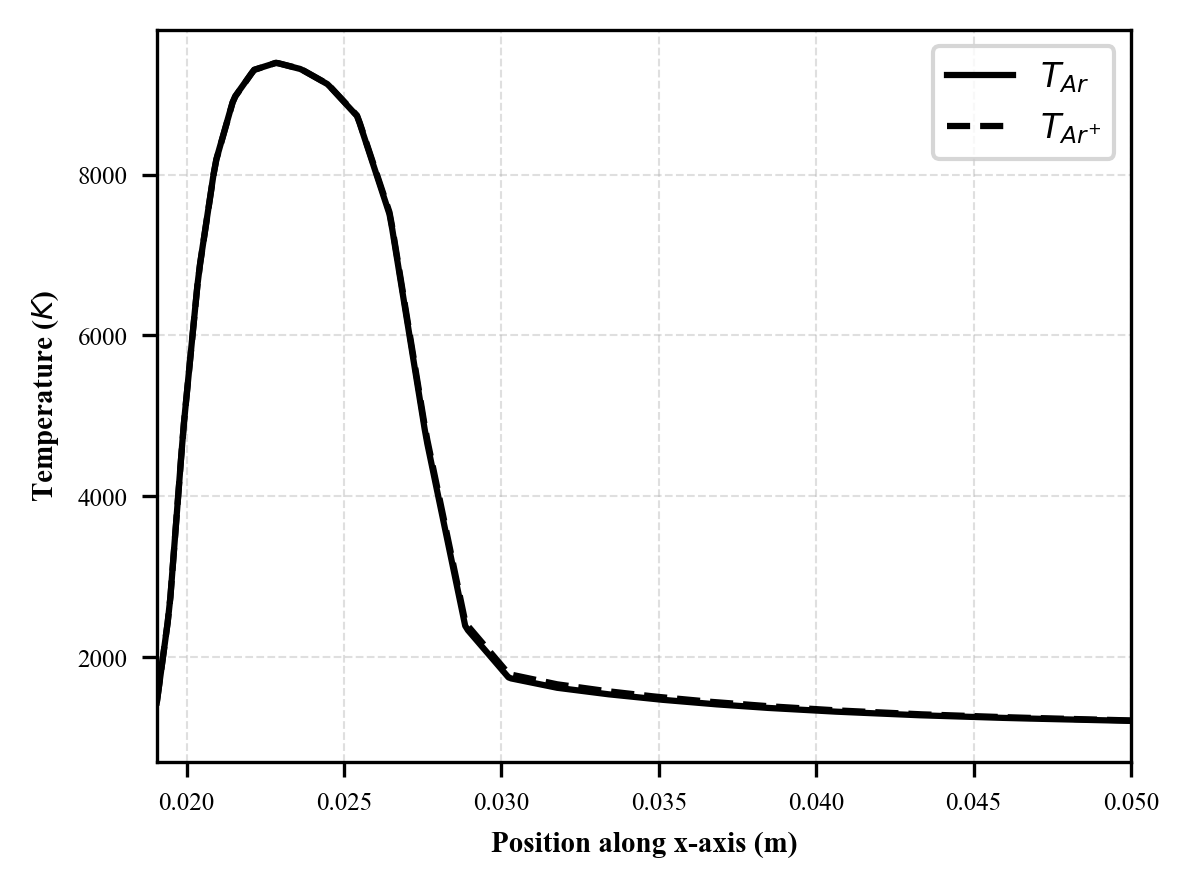}
        \caption{}
        \label{fig:Kn0.008T}
    \end{subfigure}
    \hfill 
    \caption{Temperature profile of argon atom and argon ions along the stagnation line at (a) Kn=0.2 and (b) Kn=0.008.}
    \label{fig:Tslip}
\end{figure}

Figure \ref{fig:different_Kn_contour} displays the argon atom temperature field at various Knudsen numbers. The shock front exhibits a sharpening trend as the Knudsen number decreases. The standoff distance increment reduces with increasing Knudsen numbers. As a result, the total heat flux decrement also exhibits a decreasing trend with larger Knudsen numbers as shown in Table \ref{tab:heatflux decrement}. This phenomenon is due to the enhanced decoupling between atoms and charged particles at higher Knudsen numbers. As illustrated in Figure \ref{fig:Tslip}, the left panel shows the temperature profiles of argon atoms and ions under an external magnetic field at Kn = 0.2. In this regime, the low collision frequency permits significant temperature decoupling between neutral argon atoms and ions. In contrast, the right panel presents results at Kn=0.008, where the elevated collision frequency drives the temperatures of argon atoms and ions towards equilibrium. In summary, increasing the Knudsen number amplifies the decoupling between charged and neutral particles, thereby weakening the influence of electromagnetic control effects.

\begin{table}[h!]
    \centering
    \begin{tabular}{lc}
        \hline
        \hline
        Knudsen number & Heat flux decrement\\
        \hline
         0.2 & 7.09\%  \\
        0.044 & 12.37\% \\
        0.008 & 13.71\% \\
        \hline
    \end{tabular}
    \caption{Heat flux decrement with varying Knudsen number.}
    \label{tab:heatflux decrement}
\end{table}

\section{Conclusions}
\label{sConclusions}

This study applies the UGKWP method to multiscale simulations of partially ionized plasma flows around a hemisphere, spanning regimes from near-continuum to highly rarefied. Validation against established UGKS solutions for neutral flows, along with experimental data for pre-ionized argon flows, confirms the accuracy and robustness of the numerical implementation. A comparative investigation of control effects across different Knudsen numbers shows that rarefied effects can significantly alter predictions of electromagnetic flow control. Future work will extend the framework to include additional physical phenomena, such as wall sheath dynamics and comprehensive magnetohydrodynamic effects at moderate to high magnetic Reynolds numbers. The reliability of these advanced simulations will continue to rely on the robust multiscale flow-field solutions enabled by the UGKWP solver. Taken together, the present study establishes a solid foundation for advancing the simulation of complex plasma flow phenomena.

\section{Acknowledgement}
This current work was supported by National Key R$\&$D Program of China (Grant Nos. 2022YFA1004500),
National Natural Science Foundation of China (12172316, 92371107),
and the Hong Kong Research Grant Council (16301222, 16208324).

\appendix

\section{Artificial electron mass effect}
\label{increaseB}

According to the Drude's  conductivity model \cite{yoshino2010numerical}, $$\sigma^{cond} = \frac{n_e e^2}{m_e\sum_{s\neq e}\nu_{e,s}},$$
where $\nu_{e,s}$ is the collision frequency between species $s$ and electrons and is given as \cite{chen1984introduction},
$$
\nu_{e,s} = n_s \sigma_{e,s}^{col} \bar{v}_{th}
$$
where $n_s$ is number density of species $s$, $\sigma_{e,s}^{col}$ is the cross section between species $s$ and electrons. $\bar{v}_{th}$ is the average relative velocity between species $s$ and electrons and is given approximately as the average velocities of electrons $\bar{v}_{th}=\sqrt{\frac{8k_B T_e}{\pi m_e}}$. Therefore, the conductivity is given as
\begin{equation}
\sigma^{cond} = \frac{n_e e^2}{m_e\sum_{s\neq e}n_s \sigma_{e,s}^{col}\sqrt{\frac{8k_B T_e}{\pi m_e}}}.
\label{eq:conductivity_experssion}
\end{equation}
In the current work, $\sigma_{e,s}^{col}$ is given by a fitted curve, and $n_e, T_e$ and $n_s$ are calculated by the flow field so their expressions don't contain $m_e$, then we have
$$
\sigma^{cond} \propto \frac{1}{\sqrt{m_e}},
$$
which means conductivity decreases as $m_e$ increases.
Standoff distance is proportional to the dimensionless magnetic interaction parameter \cite{poggie2002},
\begin{equation}
Q = \frac{\sigma^{cond}_0 |\boldsymbol{B}_0|^2 L_0}{\rho_0 |\boldsymbol{U}|_0},
\label{eq:interaction_parameter}
\end{equation}
where subscript $0$ means characteristic quantities. To make $Q$ constant, $|\boldsymbol{B}_0|^2$ has to increase when the conductivity decreases. Given that the actual mass ratio of argon ions ($Ar^+$) to electrons ($e^-$) is approximately 72811, while the ratio employed in this paper is 10, the electrical conductivity is consequently reduced by a factor of approximately 85, $|\boldsymbol{B}_0|^2$ needs to be increased by a factor of 85 to maintain the desired similitude.

\section{Discretizations of cross-species collision term}
\label{aap-implicit}

The discretized form of Eq.\ref{eq:aap-moment} is
\begin{align*}
    &\frac{\boldsymbol{U}_\alpha^{n+1} - \boldsymbol{U}_\alpha^{**}}{\Delta t} = \sum_{k=1}^N \frac{2 m_k \chi_{\alpha k} n_k}{m_\alpha + m_k} (\boldsymbol{U}_k^{n+1} - \boldsymbol{U}_\alpha^{n+1}),\\
    &\frac{E_\alpha^{n+1} - E_\alpha^{**}}{\Delta t} = \sum_{k=1}^N \frac{2m_{k}\chi_{\alpha k} n_k}{m_{\alpha}+m_{k}} \boldsymbol{U}_{\alpha}^{n+1}\left(\boldsymbol{U}_k^{n+1}-\boldsymbol{U}_\alpha^{n+1}\right)\\
    &\;+\sum_{k=1}^N \frac{4 m_k \chi_{\alpha k} n_k}{(m_\alpha + m_k)^2} \left[ m_k (E_k^{n+1} - \frac{1}{2}(\boldsymbol{U}_k^{n+1})^2) - m_\alpha (E_\alpha^{n+1} - \frac{1}{2}(\boldsymbol{U}_\alpha^{n+1})^2) + \frac{m_k}{2} (\boldsymbol{U}_k^{n+1} - \boldsymbol{U}_\alpha^{n+1})^2 \right].
\end{align*}
The Jacobian of residuals to velocity and energy is
\begin{align*}
    & \frac{\partial R_{U_\alpha}}{\partial \boldsymbol{U}_k} = \delta_{\alpha k}(\frac{1}{\Delta t} + \sum_{k=1}^N \frac{2 m_k \chi_{\alpha k} n_k}{m_\alpha + m_k}) - (1-\delta_{\alpha k})(\frac{2 m_k \chi_{\alpha k} n_k}{m_\alpha + m_k}),\\
    & \frac{\partial R_{E_\alpha}}{\partial E_k} = \delta_{\alpha k}(\frac{1}{\Delta t} +  \sum_{k=1}^N \frac{4 m_k m_\alpha \chi_{\alpha k} n_k}{(m_\alpha + m_k)^2}) - (1-\delta_{\alpha k}) \frac{4 m_k^2 \chi_{\alpha k} n_k}{(m_\alpha + m_k)^2}, \\
    &\frac{\partial R_{E_\alpha}}{\partial \boldsymbol{U}_k} = \delta_{\alpha k}\sum_{k=1}^N \frac{4 m_k \chi_{\alpha k} n_k}{(m_\alpha + m_k)^2} \left( -m_\alpha \boldsymbol{U}_\alpha + m_k (\boldsymbol{U}_k - \boldsymbol{U}_\alpha) \right)+\\&\quad(1-\delta_{\alpha k})(\frac{4 m_k^2 \chi_{\alpha k} n_k}{(m_\alpha + m_k)^2} \left( \boldsymbol{U}_k - m_k (\boldsymbol{U}_k - \boldsymbol{U}_\alpha) \right))
\end{align*}
where 
$$
\delta_{\alpha k} = \left\{ 
\begin{array}{cc}
    1 &  \text{if} \alpha = k\\
    0 &  \text{else}
\end{array}
\right.
$$
Residuals are
\begin{align*}
    &R_{U_\alpha} = \frac{\boldsymbol{U}_\alpha^{n+1} - \boldsymbol{U}_\alpha^{**}}{\Delta t} - \sum_{k=1}^N \frac{2 m_k \chi_{\alpha k} n_k}{m_\alpha + m_k} (\boldsymbol{U}_k^{n+1} - \boldsymbol{U}_\alpha^{n+1}),\\
    & R_{E_\alpha} = \frac{E_\alpha^{n+1} - E_\alpha^{**}}{\Delta t} - \sum_{k=1}^N \frac{2m_{k}\chi_{\alpha k} n_k}{m_{\alpha}+m_{k}} \boldsymbol{U}_{\alpha}^{n+1}\left(\boldsymbol{U}_k^{n+1}-\boldsymbol{U}_\alpha^{n+1}\right)\\
    &\;+\sum_{k=1}^N \frac{4 m_k \chi_{\alpha k} n_k}{(m_\alpha + m_k)^2} \left[ m_k (E_k^{n+1} - \frac{1}{2}(\boldsymbol{U}_k^{n+1})^2) - m_\alpha (E_\alpha^{n+1} - \frac{1}{2}(\boldsymbol{U}_\alpha^{n+1})^2) + \frac{m_k}{2} (\boldsymbol{U}_k^{n+1} - \boldsymbol{U}_\alpha^{n+1})^2 \right].
\end{align*}
Then iterative solvers like the Newton method can be used to solve the nonlinear system.

\bibliographystyle{elsarticle-num}
\bibliography{ref}

\begin{thebibliography}{10}
\expandafter\ifx\csname url\endcsname\relax
  \def\url#1{\texttt{#1}}\fi
\expandafter\ifx\csname urlprefix\endcsname\relax\def\urlprefix{URL }\fi
\expandafter\ifx\csname href\endcsname\relax
  \def\href#1#2{#2} \def\path#1{#1}\fi

\bibitem{ali2024magnetoaerodynamics}
H.~K. Ali, Magnetoaerodynamics and hypersonics for planetary exploration and national defense, in: AIAA SCITECH 2024 Forum, p. 1422.

\bibitem{resler1958}
E.~Resler~Jr, W.~Sears, The prospects for magneto-aerodynamics, Journal of the Aerospace Sciences 25~(4) (1958) 235--245.

\bibitem{bush1958}
W.~B. Bush, Magnetohydrodynamic-hypersonic flow past a blunt body, Journal of the Aerospace Sciences 25~(11) (1958) 685--690.

\bibitem{ziemer1958}
R.~W. Ziemer, W.~B. Bush, Magnetic field effects on bow shock stand-off distance, Physical Review Letters 1~(2) (1958) 58.

\bibitem{cambel1967a}
A.~B. CAMBEL, R.~W. PORTER, Hall effect in flight magnetogasdynamics., AIAA journal 5~(12) (1967) 2208--2213.

\bibitem{cambel1967}
A.~CAMBEL, Experimental investigation of magnetoaerodynamic flow around blunt bodies, in: 6th Electric Propulsion and Plasmadynamics Conference, 1967, p. 729.

\bibitem{poggie2002}
J.~Poggie, D.~V. Gaitonde, Magnetic control of flow past a blunt body: Numerical validation and exploration, Physics of Fluids 14~(5) (2002) 1720--1731.

\bibitem{fujino2008}
T.~Fujino, T.~Yoshino, M.~Ishikawa, Numerical analysis of reentry trajectory coupled with magnetohydrodynamics flow control, Journal of Spacecraft and Rockets 45~(5) (2008) 911--920.

\bibitem{kai2017thermal}
L.~Kai, L.~Jun, L.~Weiqiang, Thermal protection performance of magnetohydrodynamic heat shield system based on multipolar magnetic field, Acta Astronautica 136 (2017) 248--258.

\bibitem{fawley2022assessment}
D.~M. Fawley, Z.~R. Putnam, S.~D'Souza, A.~Borner, Assessment of electrical conductivity in rarefied flow about mars entry vehicles, in: AIAA SCITECH 2022 Forum, 2022, p. 0825.

\bibitem{katsurayama2008kinetic}
H.~Katsurayama, M.~Kawamura, A.~Matsuda, T.~Abe, Kinetic and continuum simulations of electromagnetic control of a simulated reentry flow, Journal of Spacecraft and Rockets 45~(2) (2008) 248--254.

\bibitem{parent2023}
B.~Parent, P.~T. Rajendran, S.~O. Macheret, J.~Little, R.~W. Moses, C.~O. Johnston, F.~M. Cheatwood, Effect of plasma sheaths on earth-entry magnetohydrodynamics, Journal of thermophysics and heat transfer 37~(4) (2023) 845--857.

\bibitem{lani2023}
A.~Lani, V.~Sharma, V.~F. Giangaspero, S.~Poedts, A.~Viladegut, O.~Chazot, J.~Giacomelli, J.~Oswald, A.~Behnke, A.~S. Pagan, et~al., A magnetohydrodynamic enhanced entry system for space transportation: Meesst, Journal of Space Safety Engineering 10~(1) (2023) 27--34.

\bibitem{boyd2017nonequilibrium}
I.~D. Boyd, T.~E. Schwartzentruber, Nonequilibrium gas dynamics and molecular simulation, Vol.~42, Cambridge University Press, 2017.

\bibitem{liu2020unified}
C.~Liu, Y.~Zhu, K.~Xu, Unified gas-kinetic wave-particle methods i: Continuum and rarefied gas flow, Journal of Computational Physics 401 (2020) 108977.

\bibitem{zhu2019unified}
Y.~Zhu, C.~Liu, C.~Zhong, K.~Xu, Unified gas-kinetic wave-particle methods. ii. multiscale simulation on unstructured mesh, Physics of Fluids 31~(6) (2019).

\bibitem{liu2021unified}
C.~Liu, K.~Xu, Unified gas-kinetic wave-particle methods iv: multi-species gas mixture and plasma transport, Advances in Aerodynamics 3 (2021) 1--31.

\bibitem{pu2024gas}
Z.~Pu, C.~Liu, K.~Xu, Gas-kinetic scheme for partially ionized plasma in hydrodynamic regime, Journal of Computational Physics 505 (2024) 112905.

\bibitem{yang2022unified}
X.~Yang, W.~Shyy, K.~Xu, Unified gas-kinetic wave--particle method for gas--particle two-phase flow from dilute to dense solid particle limit, Physics of Fluids 34~(2) (2022).

\bibitem{yang2024unified}
X.~Yang, W.~Shyy, K.~Xu, Unified gas-kinetic wave--particle method for polydisperse gas--solid particle multiphase flow, Journal of Fluid Mechanics 983 (2024) A37.

\bibitem{liu2023implicit}
C.~Liu, W.~Li, Y.~Wang, P.~Song, K.~Xu, An implicit unified gas-kinetic wave--particle method for radiative transport process, Physics of Fluids 35~(11) (2023).

\bibitem{yang2025unified}
X.~Yang, Y.~Zhu, C.~Liu, K.~Xu, Unified gas-kinetic wave-particle method for frequency-dependent radiation transport equation, Journal of Computational Physics 522 (2025) 113587.

\bibitem{liu2025unified}
H.~Liu, X.~Yang, C.~Zhang, X.~Ji, K.~Xu, Unified gas-kinetic wave-particle method for multi-scale phonon transport, arXiv preprint arXiv:2505.09297 (2025).

\bibitem{pu2025unified}
Z.~Pu, K.~Xu, Unified gas-kinetic wave-particle method for multiscale flow simulation of partially ionized plasma, Journal of Computational Physics 530 (2025) 113918.

\bibitem{andries2002consistent}
P.~Andries, K.~Aoki, B.~Perthame, A consistent bgk-type model for gas mixtures, Journal of Statistical Physics 106 (2002) 993--1018.

\bibitem{morse1963}
T.~Morse, Energy and momentum exchange between nonequipartition gases, The Physics of Fluids 6~(10) (1963) 1420--1427.

\bibitem{bird1976molecular}
G.~A. Bird, Molecular gas dynamics, NASA STI/Recon Technical Report A 76 (1976) 40225.

\bibitem{munz2000divergence}
C.-D. Munz, P.~Omnes, R.~Schneider, E.~Sonnendr{\"u}cker, U.~Voss, Divergence correction techniques for maxwell solvers based on a hyperbolic model, Journal of Computational Physics 161~(2) (2000) 484--511.

\bibitem{long2024implicit}
W.~Long, Y.~Wei, K.~Xu, An implicit adaptive unified gas-kinetic scheme for steady-state solutions of nonequilibrium flows, Physics of Fluids 36~(10) (2024).

\bibitem{bisek2010numerical}
N.~J. Bisek, I.~D. Boyd, J.~Poggie, Numerical study of magnetoaerodynamic flow around a hemisphere, Journal of Spacecraft and Rockets 47~(5) (2010) 816--827.

\bibitem{devoto1973transport}
R.~S. Devoto, Transport coefficients of ionized argon, Physics of Fluids 16~(5) (1973) 616--623.

\bibitem{long2024nonequilibrium}
W.~Long, Y.~Wei, K.~Xu, Nonequilibrium flow simulations using unified gas-kinetic wave-particle method, AIAA Journal 62~(4) (2024) 1411--1433.

\bibitem{yoshino2010numerical}
T.~Yoshino, T.~Fujino, M.~Ishikawa, Numerical study of thermal protection utilizing magnetohydrodynamic technology in super-orbital reentry flight, in: 41st Plasmadynamics and Lasers Conference, 2010, p. 4486.

\bibitem{chen1984introduction}
F.~F. Chen, et~al., Introduction to plasma physics and controlled fusion, Vol.~1, Springer, 1984.

\end{thebibliography}
\end{document}